\documentclass[11pt,a4paper]{article}
\pdfoutput=1

\usepackage{jheppub}
\usepackage{subfigure}
\usepackage{footnote}
\usepackage[T1]{fontenc}
\usepackage{rotating}
\usepackage{array}
\usepackage{amsmath}
\usepackage{slashed,hyperref}
\usepackage[normalem]{ulem}
\usepackage{cancel}

\newcommand{\be}{\begin{equation}}
\newcommand{\ee}{\end{equation}}
\newcommand{\ba}{\begin{eqnarray}}
\newcommand{\ea}{\end{eqnarray}}

\title{ \boldmath{Hunting electroweakinos at future hadron colliders and direct detection experiments}}
 
\author{Giovanni Grilli di Cortona}

\affiliation{\textit{SISSA - International School for Advanced Studies,\\
Via Bonomea 265, I-34136 Trieste, Italy }}
\affiliation{\textit{INFN, Sezione di Trieste,\\
via Valerio 2, I-34127 Trieste, Italy}}

\emailAdd{ggrilli@sissa.it}
\abstract{We analyse the mass reach for electroweakinos at future hadron colliders and their interplay with direct detection experiments. Motivated by the LHC data, we focus on split supersymmetry models with different electroweakino spectra. We find for example that a 100 TeV collider may explore Winos up to $\sim 7$ TeV in low scale gauge mediation models or thermal Wino dark matter around $3$ TeV in models of anomaly mediation with long-lived Winos. We show moreover how collider searches and direct detection experiments have the potential to cover large part of the parameter space even in scenarios where the lightest neutralino does not contribute to the whole dark matter relic density.}
 \arxivnumber{ }
\begin{document}
\maketitle
\flushbottom

\section{Introduction}

The LHC is getting ready to start its second run at a centre of mass energy of $13$ TeV in 2015. In the first run ATLAS and CMS have discovered the Higgs boson \cite{Aad:2012tfa,Chatrchyan:2012ufa} and have put a huge effort into looking for new physics. Unfortunately new phenomena pointing to beyond the standard model scenarios did not show up. However, this fact is not discouraging: new physics may be within the reach of LHC or future colliders.   

The lack of discovery of new coloured states at the LHC might be interpreted as in tension with a natural implementation of supersymmetric (SUSY) models (see for example \cite{Craig:2013cxa}), favouring instead split SUSY models \cite{Wells:2003tf,ArkaniHamed:2004fb,Giudice:2004tc,ArkaniHamed:2004yi}. In split SUSY scalars are heavy and the low energy spectrum is characterised by gauginos and higgsinos. Split SUSY models give up on the idea that SUSY solves the hierarchy problem but it ameliorates other potential  problems of ordinary SUSY such as flavour changing neutral currents, CP violation and fast proton decay. Moreover split scenarios maintain the successful unification of the gauge couplings and the lightest supersymmetric particle (LSP) as a viable candidate for dark matter. Indeed, if the lightest neutralino is stable, it may play the role of dark matter candidate. Pure higgsino ($\tilde{h}$) or Wino ($\tilde{W}$) states provide the full dark matter abundance if their masses are $1.1$ and $3.2$ TeV, respectively. A pure Bino ($\tilde{B}$) is not a suitable candidate to be the whole dark matter because it would be overproduced in the early universe. However, there are other suitable scenarios in which the dark matter particle is an admixture of two states: Bino/higgsino ($\tilde{B}/\tilde{h}$), Bino/Wino ($\tilde{B}/\tilde{W}$) and Wino/higgsino ($\tilde{W}/\tilde{h}$) \cite{Giudice:2004tc,ArkaniHamed:2004yi,Pierce:2004mk,Masiero:2004ft,Baer:2005zc,Baer:2005jq,BirkedalHansen:2001is,BirkedalHansen:2002am,ArkaniHamed:2006mb}. 

The scientific community already started discussing collider physics beyond the LHC: in particular there are plans for hadron colliders with centre of mass energy up to $100$ TeV.
Indeed, in several SUSY scenarios, the Higgs mass and the lack of discovery of new coloured particles give a hint for a new physics scale higher than the electroweak scale. Moreover, in SUSY models, thermal dark matter candidates might have a mass up to $3$ TeV, not accessible at the LHC. The expectation of such rather massive SUSY spectrum would inevitably require a future more powerful machine than the LHC. The exploration of higher energy through 33 TeV or 100 TeV hadron colliders seems thus to be necessary.
The aim of this paper is, therefore, to study the physics capability of future hadron colliders and to show the complementary roles of collider searches and direct detection experiments in several SUSY scenarios.

Some simplified models where scalars are heavy and electroweakinos are the only accessible states were studied at the LHC by ATLAS and CMS \cite{Aad:2014nua,Khachatryan:2014qwa}.\footnote{A reinterpretation of the LHC results for light neutralino dark matter can be found for example in \cite{Calibbi:2013poa,Calibbi:2014lga}.} In particular cases they also studied the projection for the 14 TeV LHC \cite{ATLAS-PUB-2014-010,CMS-PAS-FTR-13-2014}.

The WIMP (weakly interacting massive particle) nature of the lightest neutralino can be explored in collider searches, direct \cite{Xiao:2014xyn,Agnese:2014aze,Akerib:2013tjd,Aprile:2012nq,Aprile:2013doa} and indirect \cite{2011PhRvL.107x1302A,2010PhRvL.105l1101A} detection experiments. In this paper we focus only on the first two types of searches because the latter is dominated by large astrophysical uncertainties (see for example \cite{Cohen:2013ama,Fan:2013faa,Hryczuk:2014hpa}).  Recently there has been a lot of effort in determining what could be the mass reach for the next generation colliders ($100$ TeV or so). Mass reach for coloured sparticles were analysed in simplified models assuming a $100$ TeV collider \cite{Cohen:2013xda,Cohen:2014hxa}. Pure Winos or higgsinos were studied in the mono-jet \cite{Zhou:2013raa} and mono-photon, soft lepton and long lived particles searches \cite{Low:2014cba,Cirelli:2014dsa}. When this work was near completion \cite{Acharya:2014pua,Gori:2014oua} came out with some overlap to this study. We discuss in the text differences and similarities whenever relevant. Direct detection for SUSY models was studied in the past (see for example \cite{Masiero:2004ft,Baer:2005zc,Baer:2005jq,BirkedalHansen:2001is,BirkedalHansen:2002am,Baltz:2004aw}) and recently looking for blind spots \cite{Cheung:2012qy,Han:2013gba,Cheung:2013dua,Huang:2014xua,Han:2014nba}.

The paper is organised in two parts. In the first part  (section \ref{sec:futurereach}) we introduce the method used for the determination of the future reach at hadron colliders and describe the searches analysed. The scenarios considered were chosen because of the presence of the experimental analyses and they are meaningful and motivated examples of split SUSY models. The chosen set of options suffice out of a broad range of possibilities and it does not exhaust them. We analysed the following simplified models (the details of each simplified model are spelled out in the corresponding section):
\begin{itemize}
\item gravity or gauge mediation models with Bino LSP and Wino NLSP, where the charged Wino decays to $W$ and LSP and the neutral Wino decays either to $Z$ and LSP or to Higgs and LSP (section \ref{sec:Wino-Bino}) \cite{Aad:2014nua,Khachatryan:2014qwa,ATLAS2013-093}. The ATLAS and CMS experiment already exclude a Wino NLSP for $m\lesssim350$ GeV for Bino masses less than 100 GeV in  the \emph{WZ} channel or $m\lesssim287$ GeV in the \emph{Wh} channel (leaving unexplored a window between $141$ and $166$ GeV);
\item anomaly mediated models with long lived Winos (section \ref{sec:llWino}). In this simplified model the LSP and NLSP are almost degenerate Winos. ATLAS excludes charginos with a mass below 250 GeV in AMSB \cite{Aad:2013yna}. This search has crucial importance in the hunt for dark matter since a Wino LSP  is expected to thermally saturate the relic density for $m\sim3.2$ TeV;
\item low scale gauge mediation models with universal gaugino masses with Bino NLSP decaying to photon and gravitino (section \ref{sec:GMWino}). ATLAS set limits on a Wino mass of $570$ GeV for a bino above $50$ GeV \cite{ATLAS-2014-001};
\item low scale gauge mediation models with higgsino NLSP decaying to gravitino and Z or W bosons (section \ref{sec:GMhiggsino}). CMS excludes higgsinos with a mass less than $350$ GeV in this scenario  \cite{Khachatryan:2014qwa}.
\end{itemize}
 In the second part we show the current bounds and future reach from direct detection experiments for split SUSY models with universal gaugino masses (in section \ref{sec:UGM}) and models of anomaly mediation (in section \ref{sec:AMSBmodel}). In those sections we also compare collider searches with direct detection experiments for the models studied.

\section{Future reach at hadron colliders}
\label{sec:futurereach}

In this section we will extrapolate the mass reach for future hadron colliders for several searches on electroweakinos relevant for split
 SUSY.  In general it is quite difficult to estimate the mass reach for future colliders because cuts, acceptances ($a$), efficiencies $(\epsilon)$ and type of analyses change and because of our ignorance on the details of the detector. In the following we will assume that cuts can be rescaled such that efficiencies and acceptances can be kept constant $(\epsilon a \simeq const)$. We basically follow the same strategy outlined in Collider Reach \cite{CR,CR2}. The energy dependence of the number of signal and background events ($S$ and $B$ respectively) is thus determined by the production cross section. In particular the energy dependence of $S$ and $B$ is the same since the parton level cross section has the same scaling $\sigma\sim 1/E^2$ at high energies and the pdf of both signal and background are evaluated at the same energy.\footnote{This is true away from the squeezed limit, where most of the background come from softer SM particles. For this reason we will restrict to the case $m_{LSP}\ll m_{NLSP}$.} Therefore, requiring that the significance at the new collider is the same as the one setting the current bounds gives
 \be
 \sigma=\frac{S}{\sqrt{B}}=\frac{S'}{\sqrt{B'}} \qquad \Rightarrow \qquad \frac{S}{S'}=1,
 \ee
 where $S'$ and $B'$ refer to the number of signal and background events at a future collider.
 
Given an existing LHC bound, the corresponding mass reach at the new collider can thus be obtained by simply computing the production cross section and requiring the same number of signal events needed to put the original bound. Since in the ratio $S'/S$ the main NLO effects cancel, the number of signal events is computed using the cross section of electroweakinos at leading order \cite{Debove:2008nr} convoluted\footnote{Computing the signal, the cross section can be factorized out from the convolution with the Parton Distribution Function $S\sim \sigma_\chi \cdot pdf$ because the integral is dominated only by the threshold $\tau_0\sim4m^2$. We verified numerically the negligible effects of the tail of the distribution. } with the MSTW Parton Distribution Function \cite{Martin:2009iq}. 

We will show in the following that our analysis on electroweakino searches is in agreement with existing studies in the literature when available. Results are shown in figures \ref{fig:WZ}--\ref{fig:GMSBhiggsino} and in table \ref{tab:collider} and refer to $95\%$ CL mass reach.


\subsection{Wino-Bino simplified model}
\label{sec:Wino-Bino}
\begin{figure} [t]
\begin{center} 
\includegraphics[scale=0.90] {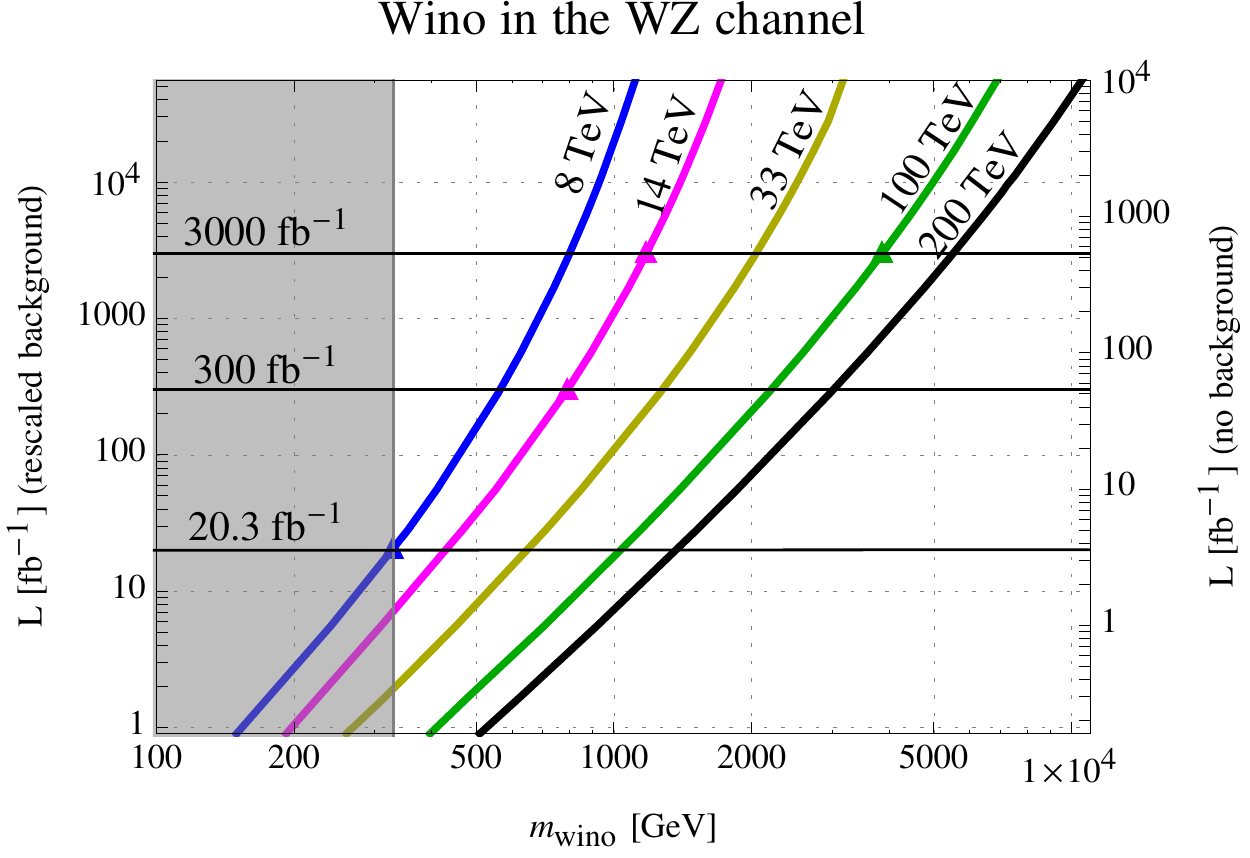}
\end{center}
\caption{Wino-Bino simplified model in the \emph{WZ} channel. The left axis shows the integrated luminosity for the method explained in the text and on the right axis, the same for having $5$ events and no background. The grey shaded area is the current bound from \cite{Aad:2014nua}.} 
\label{fig:WZ} 
\end{figure}
The first search we consider is a Wino-Bino simplified model.\footnote{The Wino-higgsino simplified model has been recently studied in \cite{Acharya:2014pua}.} Charginos can be produced in association with a neutralino via an $s$-channel $W$ boson. Production through squarks has been neglected because all the scalar super partners are assumed much heavier.
This scenario can be realised both in gravity and in gauge mediation (GMSB) models. When the gaugino masses are universal ($M_1:M_2:M_3=\alpha_1:\alpha_2:\alpha_3$), the gluino is only three times heavier than the Wino and we expect direct gluino searches to be stronger than direct Wino searches.  However, in non-universal gaugino models the gluino can be much heavier than the Wino and direct electroweak searches would be the best channel to explore this scenario.
This channel is also sensitive to GMSB models where the Wino is the lightest neutralino, all the other gauginos are heavy and the gravitino is approximately massless. In this case Winos decay promptly through the same channel $\tilde{W}^{\pm}\tilde{W}^{0}\to W^\pm \,Z \,\tilde{G}\,\tilde{G}$. 

We consider the two different extreme cases, where the neutral Wino decays with $BR=1$ either to $Z$ and LSP or to Higgs and LSP. The charged Wino decays always to $W$ and LSP. In the first case the dominant signature is three leptons and missing energy and the main background comes from the SM \emph{WZ} production. A Wino NLSP for $m\lesssim 350$ GeV is excluded for Bino masses less than $100$ GeV \cite{Aad:2014nua,Khachatryan:2014qwa}. 
In the second case the final states are one charged lepton (electron or muon), missing transverse energy (from the LSP and the neutrino coming from the $W$ decay) and two $b$-jets (from the Higgs). For a massless Bino, Winos between $125$ and $141$ GeV and between $166$ and $287$ GeV are excluded \cite{ATLAS2013-093}.

\begin{figure} [t]
\begin{center} 
\includegraphics[scale=0.90] {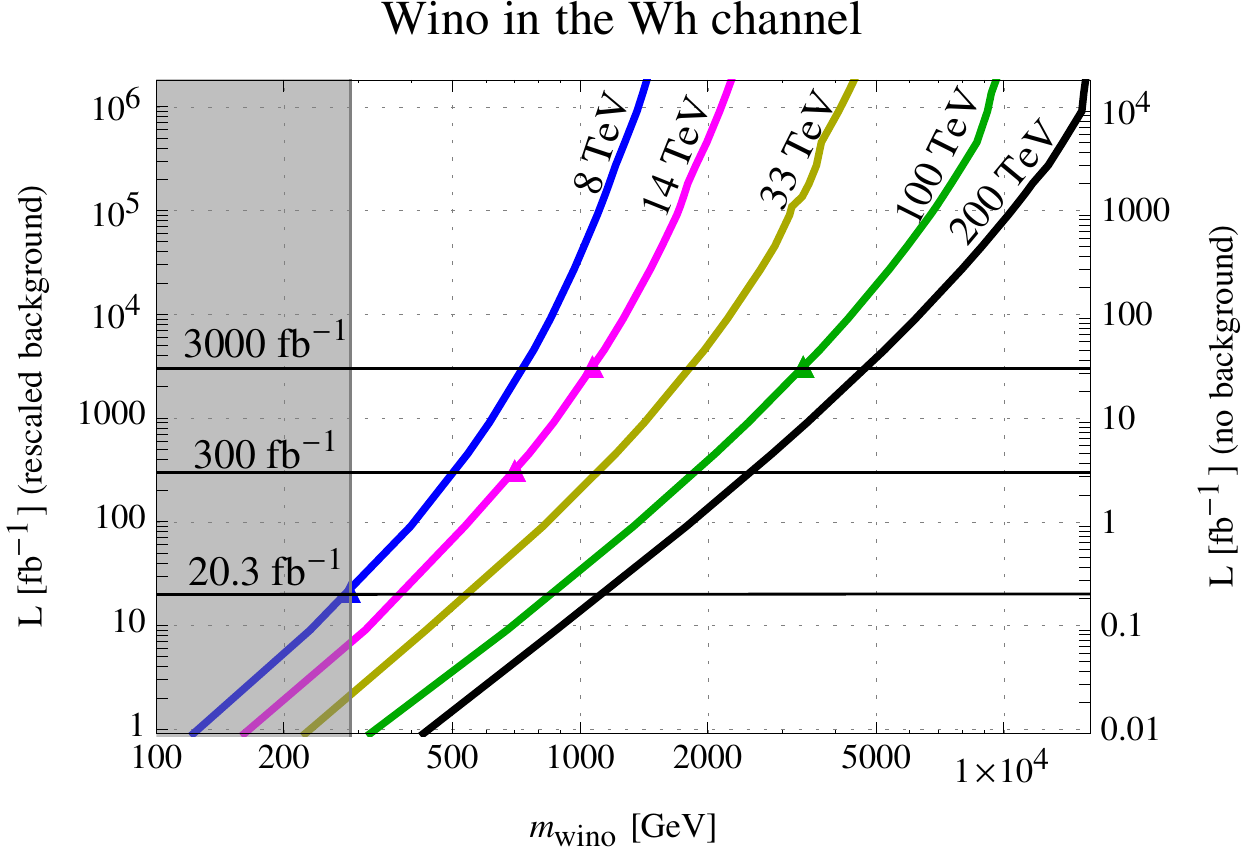}
\end{center}
\caption{Wino-Bino simplified model in the $W$ $h$ channel. The left axis shows the integrated luminosity for the method explained in the text and on the right axis, the same for having $5$ events and no background. The grey shaded area is the current bound from \cite{ATLAS2013-093}.} 
\label{fig:Wh} 
\end{figure}

The mass reaches in the Wino-Bino simplified model are shown in figure \ref{fig:WZ} and figure \ref{fig:Wh} for the \emph{WZ} and the \emph{Wh} channel respectively. In this scenario higgsinos are decoupled and therefore the only relevant parameter in the cross section is $M_2$ (the cross section is weakly dependent on $\tan\beta$ and $M_1$, fixed to be less than $100$ GeV by the method used).  In the \emph{WZ} scenario, we find that the LHC14 may extend the mass reach to $1.2$ TeV for a luminosity of $3000$ fb$^{-1}$ and it will increase up to $\sim4$ TeV at a 100 TeV collider. 

The first result is in agreement with the $1.1$ TeV mass reach given by ATLAS \cite{ATLAS-PUB-2014-010}. The latter can be compared with results by \cite{Gori:2014oua}, although the two analyses differ for the treatment of the branching ratios: we assume 100\% decay in $Z$ or $h$, while they keep into account the fact that the branching ratios of Winos depend on the choice of $\tan\beta$ and on the choice of the relative sign between gauginos and higgsinos. Moreover, in \cite{Gori:2014oua} only the lepton channels have been considered, however the $b$-jet channel has a higher sensitivity in the Higgs mediated scenario, due to the enhanced branching ratio BR$(h\to b\bar{b})\gg$BR$(h\to WW/ZZ)$. So we find that a 100 TeV collider with $3000$ fb$^{-1}$ of luminosity may reach $3.4$ TeV in the $b$-jets channel as opposed to the only $1.3$ TeV reach found in \cite{Gori:2014oua} considering only the lepton channel. 
As a reference point in the right axis of the figures we show how the reach in mass can be extended for a given integrated luminosity assuming that the background can be reduced to zero and the efficiencies and the acceptances can be made $100\%$ (obtained by requiring $S'=5$). (A more realistic result can be simply obtained by rescaling the required luminosity by $\epsilon$, $a$ and the $\sqrt{B}$.) The right axis of each figure allows also to derive the cross section for the different searches at different colliders as a function of the suitable gaugino masses.  The plots can also be used to compare the performance of different colliders. For example, in figure \ref{fig:WZ} and \ref{fig:Wh}, we notice that, for the electroweak (EW) searches described in this section, the sensitivity of the LHC14 with $3000$ fb$^{-1}$ is approximately the same as a $33$ TeV collider with a luminosity ten times smaller.  


\subsection{Long-lived Wino}
\label{sec:llWino}

Long-lived chargino searches can be used to probe models with Wino LSP such as anomaly mediation models (AMSB) or high scale GMSB with non universal gaugino masses. In these models the neutral Wino states are highly degenerate with the charged Wino and all the other states are decoupled. For heavy higgsino the mass splitting at tree level is suppressed and it is dominated by the radiative generated contribution, which is around 160-170 MeV at one-loop level \cite{Cheng:1998hc,Feng:1999fu,Gherghetta:1999sw}. This small mass splitting implies that the charged Wino has a considerable lifetime (of order $c \tau = \mathcal{O}(10)$ cm) and it decays mainly into the neutral Wino and a soft charged pion. 

The signature for this search is one hard jet from initial state radiation (ISR), large missing transverse energy and a disappearing track (the chargino eventually decays to a soft not reconstructed pion). The jet must not be too close to the missing energy direction because it usually implies jet mismeasurement. Chargino pair ($\tilde{\chi}_1^{+} \tilde{\chi}_1^{-}$) and chargino neutralino ($\tilde{\chi}_1^{\pm}\tilde{\chi}_1^{0}$) associated production with initial state radiation are the relevant processes for this search. 
The relevant background originates from unidentified leptons and charged particles with high mis-reconstructed transverse momentum ($p_T$) as well as charged hadrons interacting with the inner detector. ATLAS excludes charginos with mass below $250$ GeV in the AMSB model ~\cite{Aad:2013yna}. 

\begin{figure} [t]
\begin{center} 
\includegraphics[scale=0.90] {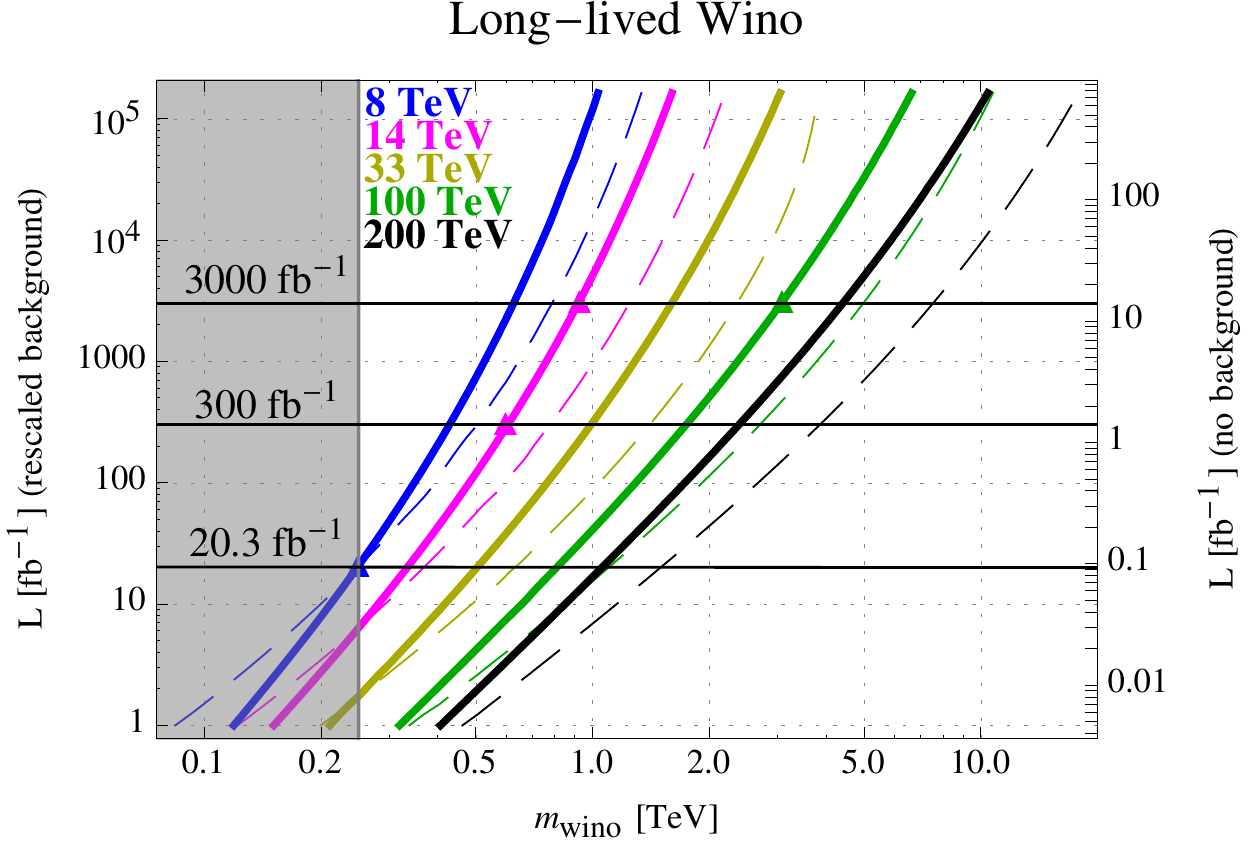}
\end{center}
\caption{Long-lived Wino. The left axis shows the integrated luminosity for the method explained in the text and on the right axis, the same for having $5$ events and no background. The grey shaded area is the current bound from \cite{Aad:2013yna}.} 
\label{fig:longlivedWino} 
\end{figure}

We model the relevant cross section through the process $q\bar{q}\to Z \, j \to e^+\,e^-\,j$ using the program MCFM \cite{Campbell:2003dd} and rescaling the partonic cross section with the electroweakino one. This is a good approximation within the method used because the cross section depends only on the energy and on the different pdfs and the process with the exchange of a photon is negligible with respect of the Z exchange diagrams. We derived the mass reach in two ways: by conservatively rescaling the cut on the transverse momentum of the jet with the mass of the final state (solid lines in figure \ref{fig:longlivedWino}), in such a way that $p_T/m_{\tilde{W}}=const$, or keeping the cut fixed to the value the ATLAS experiment used in its study ($p_T> 80$ GeV), if feasible (dashed lines in figure \ref{fig:longlivedWino}).  

This scenario is relevant for dark matter searches. Indeed a Wino LSP is expected to thermally saturate the relic density for a mass $m_\chi \simeq 3.2$ TeV.  LHC14 has the potential to explore long lived chargino scenarios for masses  around $600$ GeV for a luminosity of $300$ fb$^{-1}$. This result is in agreement with the study in \cite{Haas}. By exploiting the new tracker installed at ATLAS, the reach for this kind of search may increase up to $800$ GeV at the LHC14 with $100$ fb$^{-1}$. We find that a 100 TeV collider would reach a Wino mass around $3.1$ TeV for $3000$ fb$^{-1}$. 

In the literature there are similar results for the disappearing track of long-lived Wino searches \cite{Low:2014cba,Cirelli:2014dsa}. In order to be sure to reach the thermal dark matter mass range we should either increase the luminosity or the collider energy: for example with a 200 TeV collider and $1000$ fb$^{-1}$ of luminosity the Wino reach would comfortably extend to over $3$ TeV. In addition it seems that without stronger cuts than the one used by ATLAS the reach could be extended up to $5$ TeV for a $100$ TeV collider with a luminosity of $3000$ fb$^{-1}$. This channel is particularly important in models such as anomaly mediation, where the ratio between the gluino and the Wino is large ($M_3 \simeq\,7 \,M_2$), because it could be more powerful than the gluino searches \cite{Cohen:2013xda}.


\subsection{GMSB Wino-Bino simplified model}
\label{sec:GMWino}
\begin{figure} [t]
\begin{center} 
\includegraphics[scale=0.90] {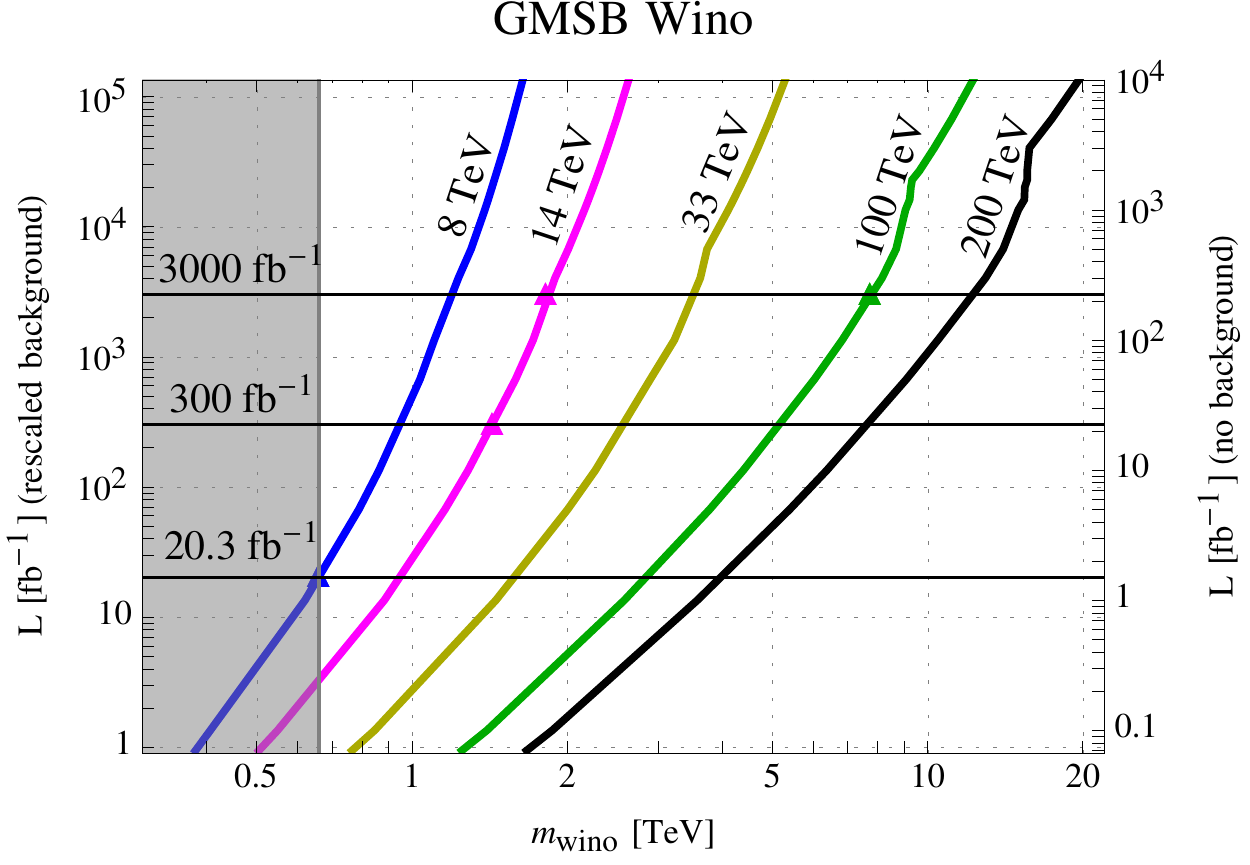}
\end{center}
\caption{GMSB Wino-Bino scenario. The left axis shows the integrated luminosity for the method explained in the text and on the right axis, the same for having $5$ events and no background. The grey shaded area is the current bound from  \cite{ATLAS-2014-001}.} 
\label{fig:GMSBwb} 
\end{figure}

In gauge mediated supersymmetric models usually the gravitino and the Bino are the LSP and the NLSP respectively and the latter decays to the former via emission of one hard photon. The search discussed in this section describes the production of Winos decaying into Binos that subsequently decay into photons and gravitinos. This channel leads to events with two final state photons, large missing energy and a moderate amount of visible transverse energy. The relevant background is given by QCD processes involving photons and jets, where a photon or a jet is mis-measured, EW processes like $W$ + $X$, where $X$ is mis-reconstructed as a photon, and $W$ and $Z$ production in association with photons. ATLAS set limits on a Wino mass of 570 GeV for any Bino above $50$ GeV \cite{ATLAS-2014-001}. For models with universal gaugino masses the limit increases to $660$ GeV.

In figure \ref{fig:GMSBwb} we show the reach for the GMSB Wino-Bino scenario with universal gaugino masses.  The relevant parameter in the computation of the cross section is the Wino mass, while higgsinos are decoupled (the cross section is weakly dependent $\tan\beta$).  Already the LHC14 will probe Winos up to $1.8$ TeV with $3000$ fb$^{-1}$, corresponding to a gluino $\sim 5.4$ TeV. At 100 TeV it is possible to exclude $\sim7.8$ TeV Wino. This has a strong impact in GMSB models with universal gaugino masses in which $\tan \beta$ is large. Indeed in these models the $\sim 125$ GeV Higgs mass fix the squark masses to be around $10$ TeV or below. Gluinos are expected at the same scale or below, which means a Wino around $\sim 3.3$ TeV or below. Such Wino could be probed already at a $33$ TeV collider with $3000$ fb$^{-1}$. Like in the previous case also in this scenario the Wino reach is stronger than the gluino one.


\subsection{GMSB higgsino simplified model}
\label{sec:GMhiggsino}

\begin{figure} [t]
\begin{center} 
\includegraphics[scale=0.90] {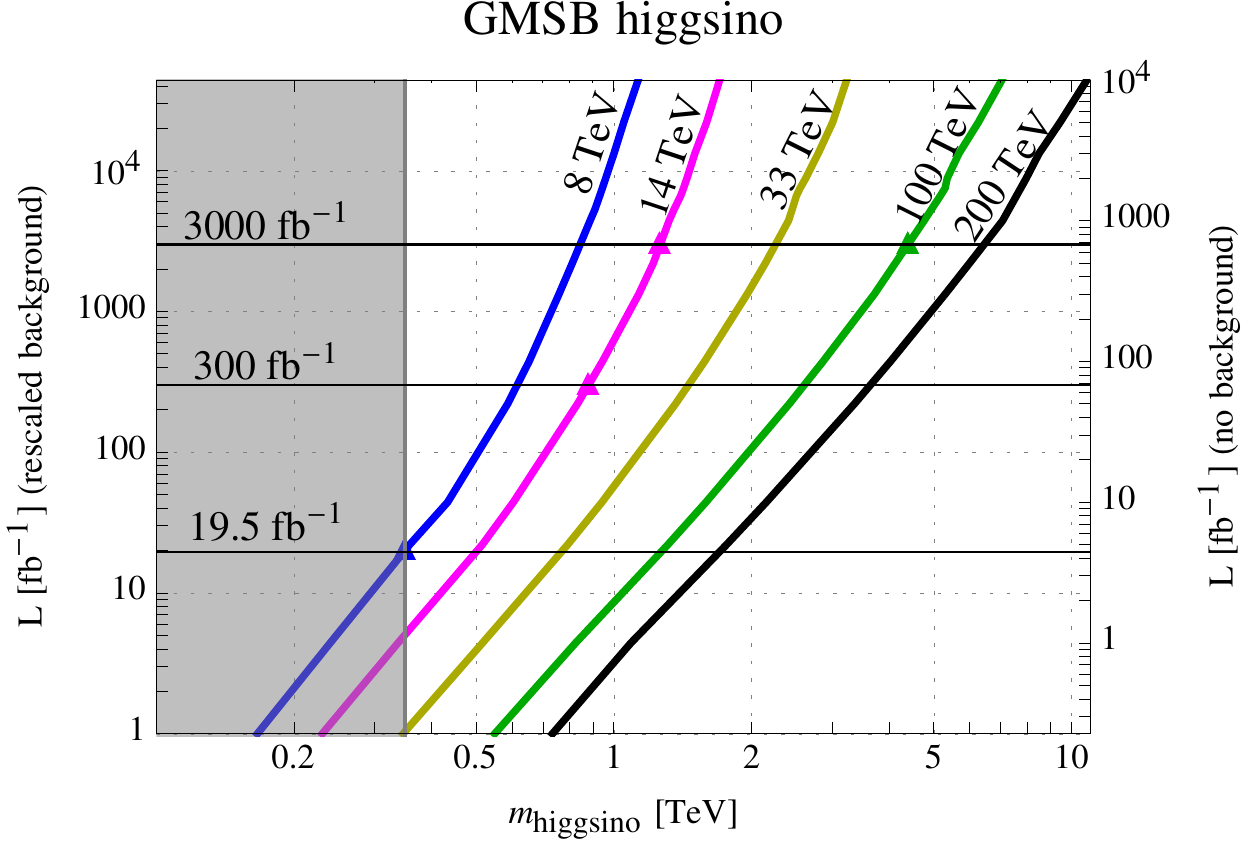}
\end{center}
\caption{GMSB higgsino NLSP scenario. The left axis shows the integrated luminosity for the method explained in the text and on the right axis, the same for having $5$ events and no background. The grey shaded area is the current bound from \cite{Khachatryan:2014qwa}.} 
\label{fig:GMSBhiggsino} 
\end{figure}

In the last analysis the gravitino is assumed to be the LSP with higgsinos NLSP and all the other states decoupled. This channel is relevant, for example, in lopsided gauge mediation models \cite{DeSimone:2011va}, where scalars and gauginos are in the multi-TeV range and the production of electroweakinos in the cascade of coloured sparticles is suppressed with respect to the direct production of light higgsinos. 

Higgsino NLSP decays to gravitino and $Z$ or $W$ bosons. The branching fraction of higgsino to $Z$ can be enhanced (with respect to the decay to Higgs) in the so called Z-enriched GMSB model \cite{Matchev:1999ft,Meade:2009qv}. The signature for this search is three or four leptons plus missing transverse momentum or two leptons, two jets and missing transverse momentum. The background is given mainly by the Standard Model \emph{WZ} and \emph{ZZ} production. CMS set a limit of $350$ GeV to this type of higgsinos  \cite{Khachatryan:2014qwa}. We consider only the channel with two leptons, two jets and missing transverse momentum since it is the one that dominates the search. 
In figure \ref{fig:GMSBhiggsino} we show the results. The mass reach of this search is expected to surpass $1$ TeV  by the end of the LHC lifetime ($\sim2035$) and reach $4.4$ TeV at a $100$ TeV machine with $3000$ fb$^{-1}$.

\begin{savenotes}
\begin{table}[t] 
\begin{center}
 \small{ 
\begin{tabular}{|c|c|c|c|c|} 
\hline
\rule[-2ex]{0mm}{5ex}%
 & $8$ TeV $^($\footnote{The limit for GMSB higgsino model was given for 19.5 fb$^{-1}$ of luminosity. All the other limits are given for $20.3$ fb$^{-1}$.}$^)$ & $14$ TeV [300(0) fb$^{-1}$]& $33$ TeV [300(0) fb$^{-1}$] & $100$ TeV [300(0) fb$^{-1}$]\\
\hline
\hline
\rule[-2ex]{0mm}{5ex}%
Wino ($\chi_2^0\to \chi_1^0 Z$) & $330$ \cite{Aad:2014nua} &  $790$ $(1180)$   & $1280$ $(2050)$ & $2210$ $(3870)$  \\
\hline
\rule[-2ex]{0mm}{5ex}%
Wino  ($\chi_2^0\to \chi_1^0 h$)& $287$  \cite{ATLAS2013-093} & $700$ $(1080)$  & $1110$ $(1830)$  &   $1890$ $(3380)$  \\
\hline
\rule[-2ex]{0mm}{5ex}%
long-lived Wino & $250$  \cite{Aad:2013yna}   & $600$ $(930)$ & $990$ $(1600)$   & $1750$ $(3080)$   \\
\hline
\rule[-2ex]{0mm}{5ex}%
GMSB Wino & $660$ \cite{ATLAS-2014-001}   & $1430$ $(1820)$ &$2590$ $(3510)$   &$5170$ $(7750)$     \\
\hline
\rule[-2ex]{0mm}{5ex}%
GMSB  higgsino& $350$ \cite{Khachatryan:2014qwa}  & $880$ $(1260)$ & $1460$ $(2260)$  & $2610$ $(4400)$   \\
\hline
\end{tabular}
}
\end{center}
\caption{Current experimental status (LHC8) and results of the analyses with rescaled background for LHC14, 33 and 100 TeV future hadron colliders. All the numbers are in GeV. The models are explained in sections \ref{sec:Wino-Bino}, \ref{sec:llWino}, \ref{sec:GMWino}, \ref{sec:GMhiggsino}  respectively. \label{tab:collider}}
\end{table}
\end{savenotes}


\section{Interplay with Direct Dark Matter searches}
\label{sec:interplay}

In split SUSY dark matter searches depend on the low energy electroweakino spectrum. We focus on two representative scenarios: models with  universal gaugino masses and models of anomaly mediated SUSY breaking. These scenarios cover all the relevant dark matter candidates in split SUSY, i.e. pure higgsino, pure Wino, $\tilde{h}/\tilde{W}$, $\tilde{h}/\tilde{B}$ and $\tilde{B}/\tilde{W}$.

\begin{figure}[t]
\centering
\subfigure[Higgs.\label{fig:Higgs}]{\includegraphics[width=0.45\textwidth]{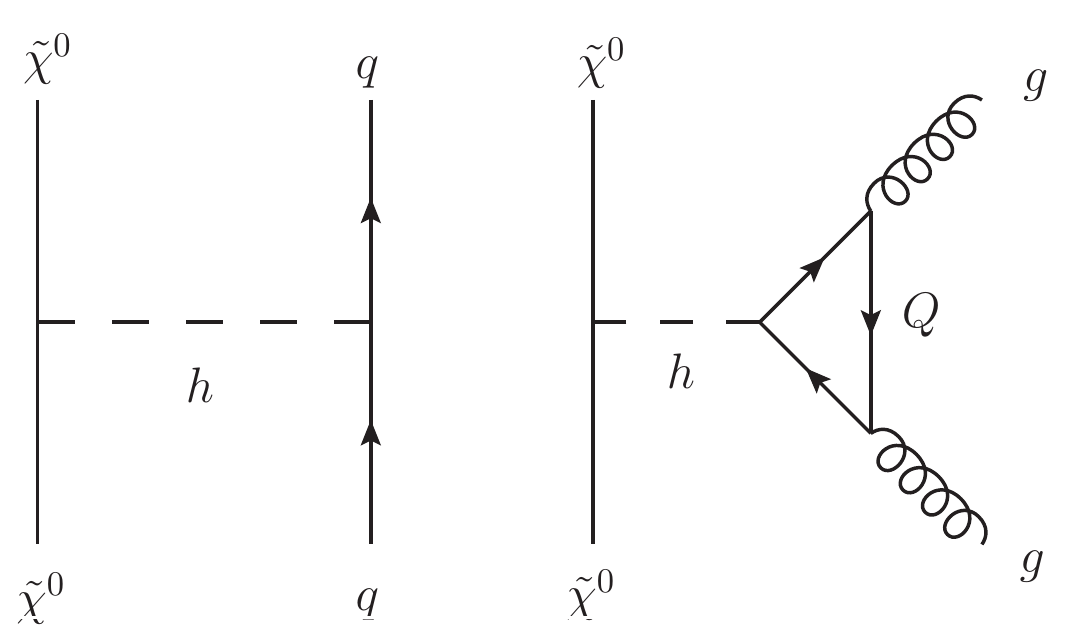}}
\hspace{1cm}
\subfigure[Twist-2.\label{fig:twist-2}]{\includegraphics[width=0.18\textwidth]{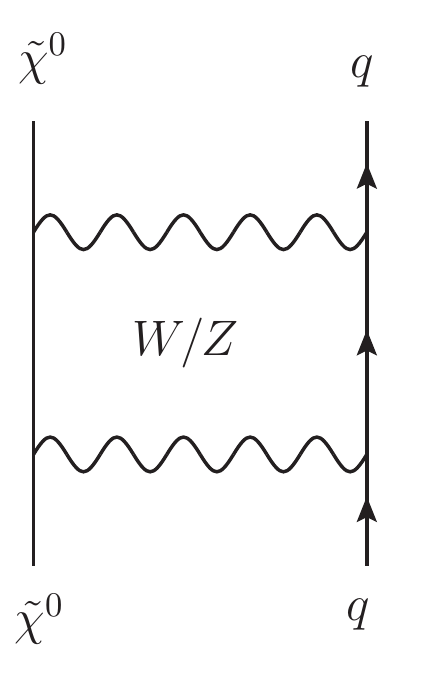}}
\subfigure[Gluon.\label{fig:gluon}]{\includegraphics[width=0.25\textwidth]{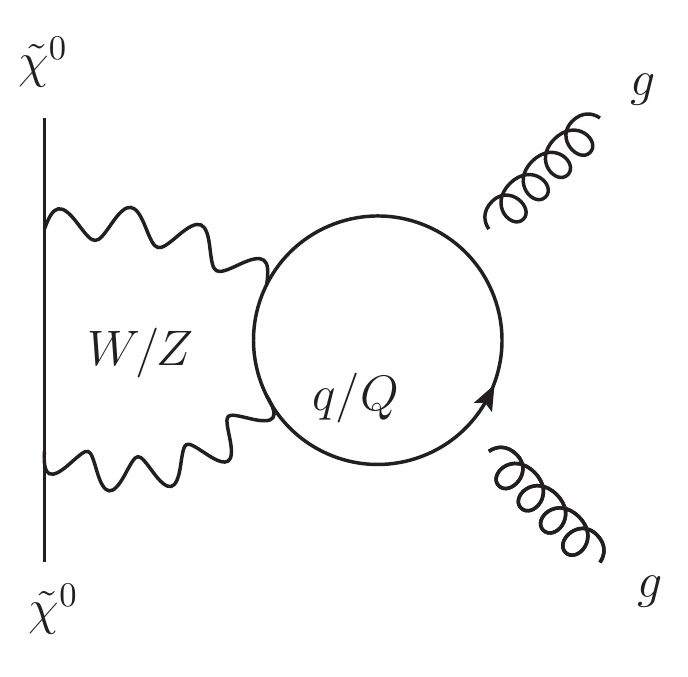}\includegraphics[width=0.25\textwidth]{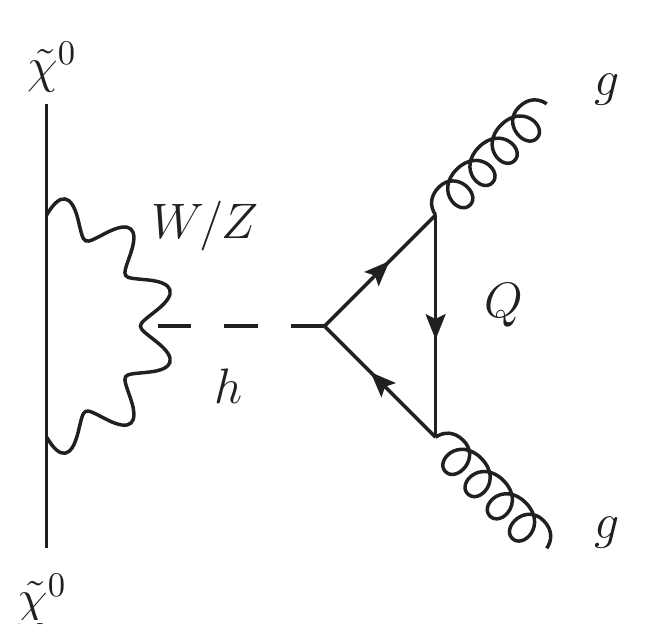}}
\caption{Diagrams that contribute to the neutralino nucleon cross section. The diagrams with $Z$ exchange are relevant only for the Spin Dependent cross section.
\label{fig:Feyn} }
\end{figure}

Direct detection experiments are based on the idea of \cite{Goodman:1984dc} that exploits the recoil energy from dark matter particles scattering on nuclei. The scattering cross section of the neutralino with nucleons is calculated using the effective Lagrangian describing the interaction among neutralinos, quarks and gluons in the limit of low relative velocity \cite{Hisano:2010fy,Hisano:2011cs,Hisano:2012wm}. The spin independent scattering cross section of the neutralino with a nucleon $N$ can be expressed in a simple way as
\be
\sigma_{N}^{SI}=|\mathrm{Higgs}+\mathrm{gluon}+\mbox{twist-2}|^2.
\ee
\emph{Higgs, twist-2} and \emph{gluon} refer to the diagrams in figure \ref{fig:Feyn}. 
The \emph{Higgs} diagrams (figure \ref{fig:Higgs}) are generated by the scalar-type effective operators $\overline{\tilde{\chi}^0}\,\tilde{\chi}^0\bar{q}q$ and $\overline{\tilde{\chi}^0}\,\tilde{\chi}^0G_{\mu\nu}^aG^{a\mu\nu}$ and their contribution to the amplitude is proportional to the $\tilde{\chi}^0$$\tilde{\chi}^0h$ coupling:
\be
c_{h\tilde{\chi}\tilde{\chi}}=(N_{12}-N_{11} \tan \theta_W)(N_{13} \cos\beta - N_{14}\sin\beta),
\label{eq:h-chi-chi}
\ee
where $N_{1i}$ are the elements of the matrix that diagonalise the neutralino mass matrix. The first subscript of $N_{1i}$ is an index in the mass basis (ordered from the lightest to the heaviest), while the second in the interaction basis $(\tilde{B},\tilde{W},H_u,H_d)$. The \emph{twist-2} diagram (figure \ref{fig:twist-2}) plays an important role in the computation of the cross section because it contributes to the amplitude with opposite sign with respect to the other diagrams. This will lead to some accidental cancellation. The \emph{gluon} contributions (figure \ref{fig:gluon}) are of the same order of the one-loop diagrams because of the presence of a factor $1/\alpha_s$ that comes from the calculation of the gluon matrix element. 

For the computation of the cross section we used leading order formulae \cite{Hisano:2010fy,Hisano:2011cs,Hisano:2012wm}. The uncertainty has been estimated by taking into account the uncertainties from hadronic matrix elements and those from known 1-loop QCD corrections. The order of magnitude of the latter is comparable with \cite{Hill:2011be,Hill:2013hoa,Hill:2014yka,Hill:2014yxa}. 

In the rest of the section we only focus on neutralino dark matter that is thermally produced. The relic density was computed with DarkSusy \cite{Gondolo:2004sc,ds} and the package DarkSE \cite{Hryczuk:2011tq} to compute the Sommerfeld effect. For the value of the relic density we used the Planck result, $\Omega h^2= 0.1196\pm0.0031$ \cite{Ade:2013zuv}.


\subsection{Models with universal gaugino masses}
\label{sec:UGM}

The first scenario we consider is split SUSY with universal gaugino masses. Higgsinos can be either light or heavy and are left as free parameters. In general we have two free parameters, $\mu$ and $M_0$ ($M_i\propto\alpha_i M_0$) which are further constrained to one by requiring $\Omega_{DM}=\Omega_{\mathrm{exp}}$. This leads to a phenomenology in which the LSP can be either the higgsino (when $|\mu|<M_1$) or the Bino (when $|\mu|>M_1$). 

\begin{figure} [t]\centering%
\begin{minipage}{0.69\linewidth}
     \includegraphics[scale=0.85] {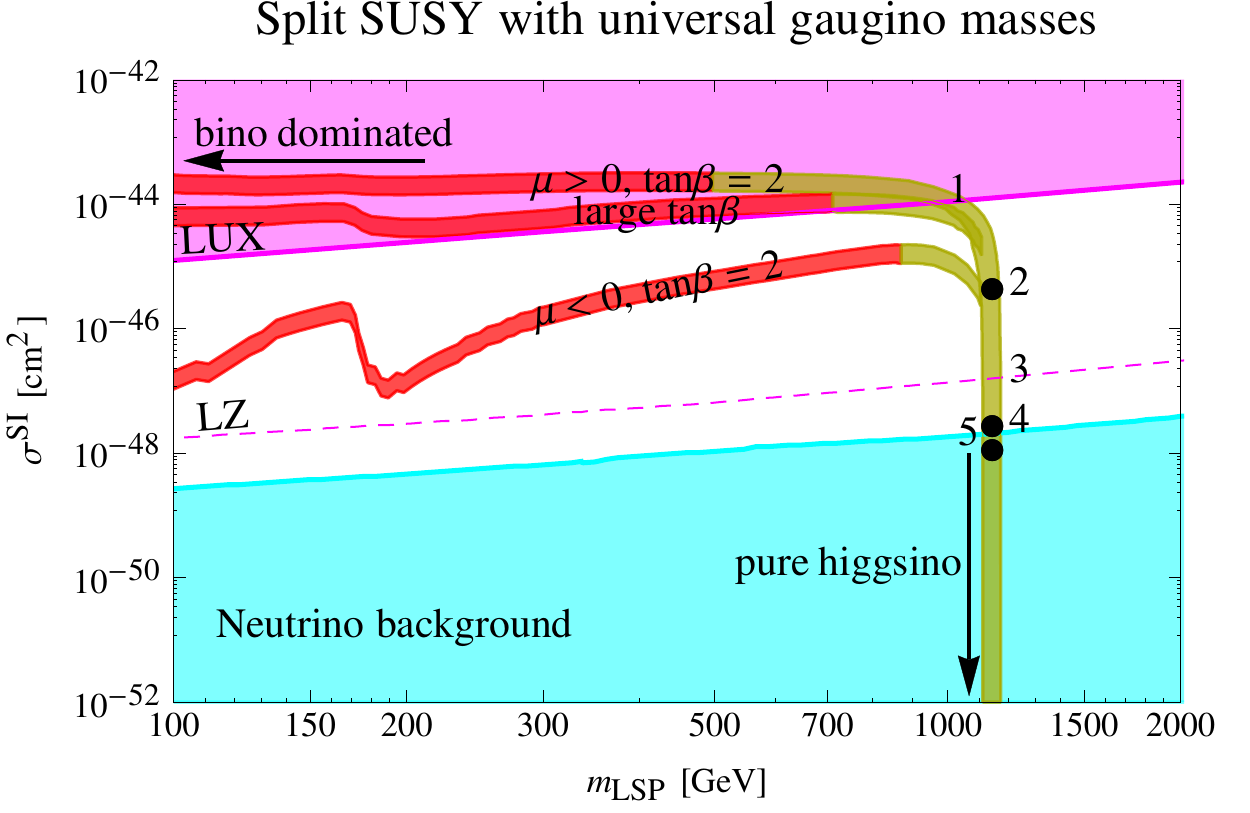}
   \end{minipage}
   \begin {minipage}{0.29\linewidth}
  \includegraphics[scale=0.5] {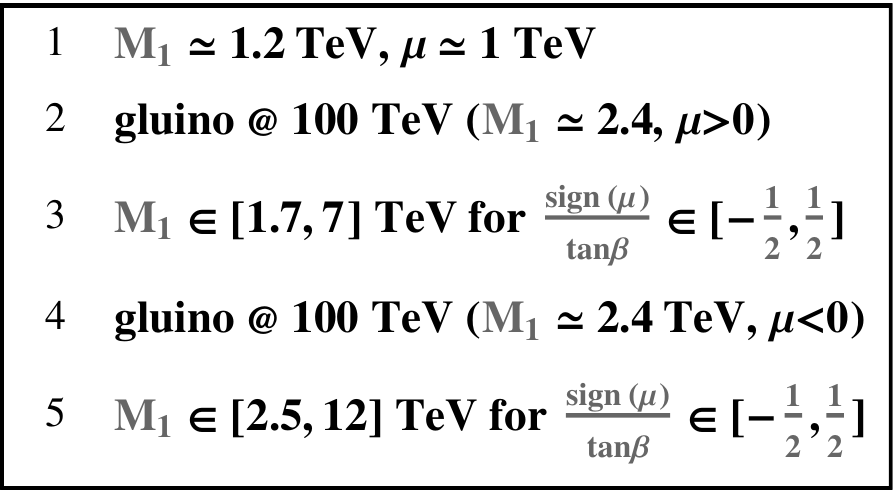}
   \end{minipage}
\caption{Spin independent neutralino-nucleon scattering cross section ($\sigma^{SI}$) requiring $\Omega h^2 =0.1196\pm0.0031$ in the universal gaugino masses model (red is Bino-like, yellow is higgsino-like LSP). The magenta area is the actual bound by LUX, the dashed magenta line is the projected reach of LZ. The neutrino background, where direct detection experiments lose sensitivity, is shaded in light blue. The numbers are explained in the legend on the right panel.} 
\label{fig:UG} 
\end{figure}

\begin{figure} [t]
\centering
\textbf{Split SUSY with universal gaugino masses}
\begin{center} 
\includegraphics[scale=0.59] {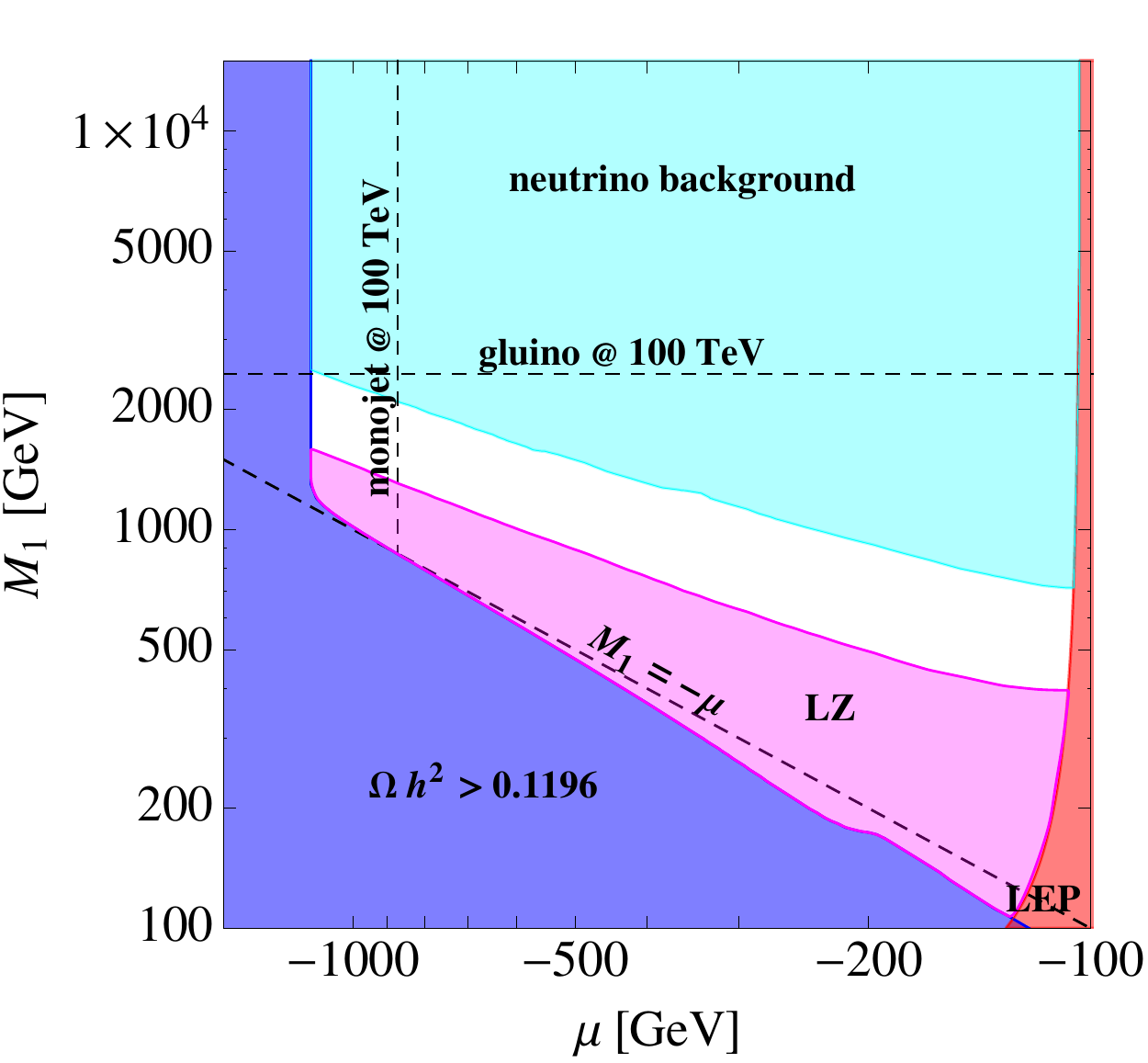}
\includegraphics[scale=0.59] {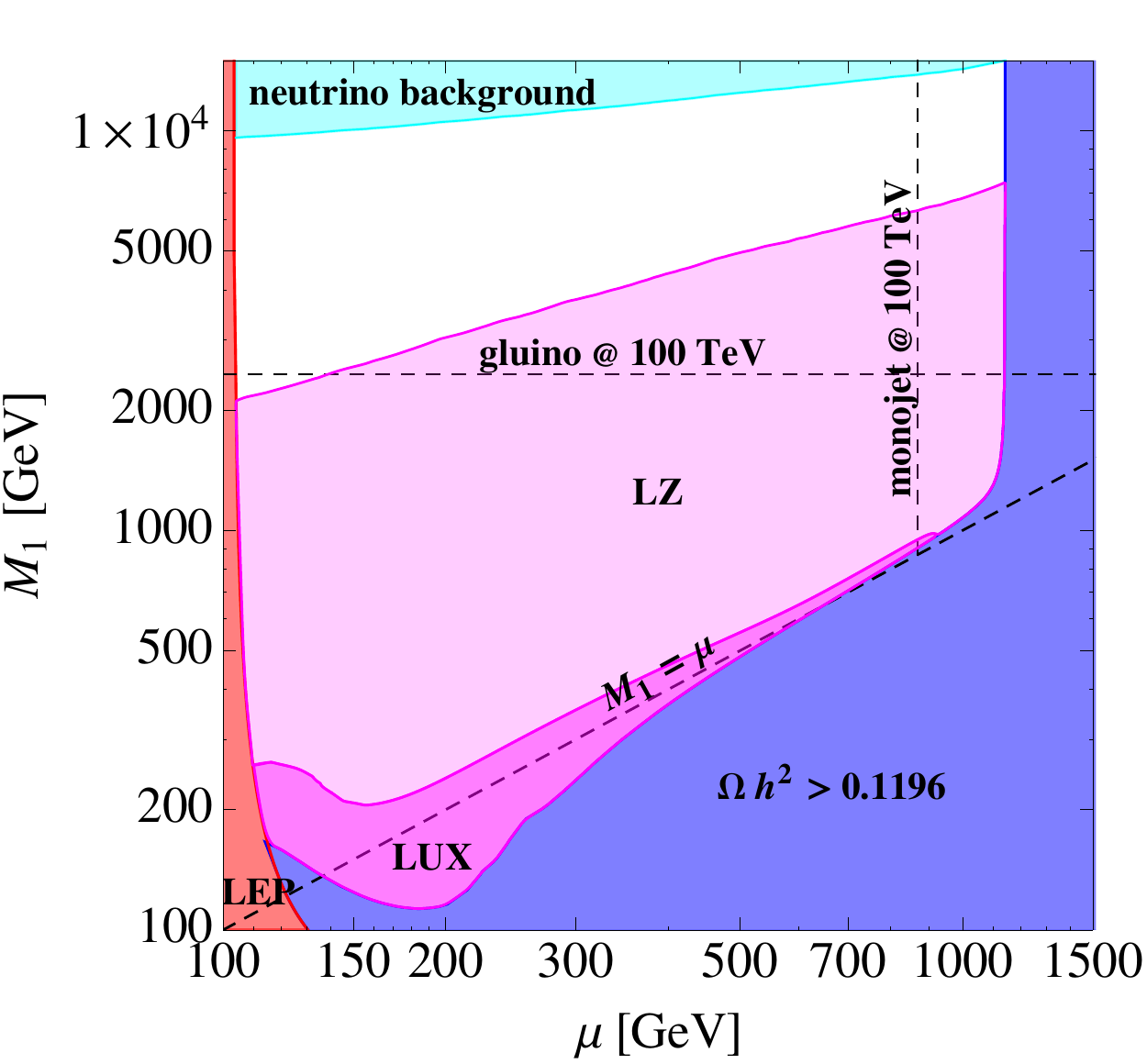}
\end{center}
\caption{Region in the ($\mu$, $M_1$) allowed by direct chargino searches at LEP (excluded in red) and by the requirement that the neutralino does not exceed the CDM relic density (excluded in blue). The left (right) panel are for negative (positive) $\mu$, respectively. The colour shading for actual bounds, future reach and background for direct detection are the same as in the previous figure.  } 
\label{fig:muvsM1} 
\end{figure}

In figure \ref{fig:UG} we show the spin independent cross section for the scattering of neutralinos on nucleons. The magenta shaded area shows the region excluded by LUX \cite{Akerib:2013tjd}. The dashed magenta curve set the projected reach for LZ \cite{LUXLZ,Cushman:2013zza}. The light blue area represents the irreducible neutrino background \cite{Billard:2013qya}. 
The three red/yellow curves represent the spin independent cross section requiring the correct relic density for $\mu>0$, large $\tan\beta$ and $\mu<0$ from top to bottom respectively (in the following we set the gaugino masses positive). For large $\tan\beta$ the sign of $\mu$ is irrelevant: the curve in the middle is the limit for both positive and negative scenario. The red colour of each curve represents a bino-like LSP, while the yellow a higgsino-like LSP. 
A Bino needs to mix with an higgsino in order to have sizeable annihilation cross section and therefore the correct relic density. In this region the cross section is dominated by the Higgs diagrams. The relic density constraint gives a relation between $\mu$ and $M_1$, depending on $\tan\beta$ and on the sign of $\mu$ (as it is shown in figures \ref{fig:muvsM1}). 
In particular for $\mu<0$ the two states are close enough and therefore coannihilation effects become relevant. At low LSP masses in the negative branch, the mixing is not maximal and it is given by
\ba
N_{11}&\simeq&1-\frac{N_{13}^2}{2}-\frac{N_{14}^2}{2}\nonumber \\
N_{13}&\simeq& -\frac{\sin \theta_\mathrm{w} M_Z}{M_1^2-\mu^2}(\mu \sin\beta+M_1 \cos\beta)\nonumber \\
N_{14}&\simeq& \frac{\sin \theta_\mathrm{w} M_Z}{M_1^2-\mu^2}(\mu \cos\beta+M_1 \sin\beta),
\label{eq:purebinomix}
\ea
while the cross section is proportional to $|M_Z (M_1 +\mu \sin2\beta)/(\mu^2-M_1^2)|^2$. As the LSP mass increases, the relic density constraint needs more mixing and more coannihilation, the two states become more degenerate and the cross section increases (figure \ref{fig:UG}). Also for $\mu>0$ at small LSP mass there is small mixing between the Bino and the higgsino. In this region coannihilation is not present.
The relic density constraint gives a relation between the parameter $\mu$ and $M_1$ such that the nucleon-neutralino scattering cross section is constant. 

Continuing the description of figure \ref{fig:UG}, the $t\bar{t}$ threshold is visible only for $\mu<0$ because for positive $\mu$ there is no coannihilation and the dominant annihilation channel is into gauge bosons.
For $m_{LSP}>m_t$ (also the annihilation in $t\bar{t}$ is present), in order to get the correct relic density the two states have to become less degenerate such that the new annihilation channel is balanced by the weaker coannihilation effect. 

Once the region of maximal mixing is reached, the $m_{LSP}\simeq 500$ (900) GeV for positive (negative) $\mu$ respectively, we have $(N_{11},N_{12},N_{13},N_{14})\simeq(1/\sqrt{2},0,1/2,\mp1/2)$, where the $\mp$ sign refers to the cases $\mu\simeq \pm M_1$. Thus the cross section is proportional to $|\cos \beta~+~ \mathrm{sign}(\mu)\sin\beta|^2$ and it is  constant for both signs of $\mu$. The suppression in the negative $\mu$ branch is again due to the sign of $\mu$ and the value of $\tan\beta$.

LUX bounds already exclude the regions $\mu<1$ TeV for small $\tan\beta$ and $|\mu| < 500$ GeV for large $\tan\beta$. When the higgsino becomes the LSP and $m_{\tilde{B}}$ become heavier and heavier, the Higgs exchange become suppressed and the \emph{twist-2} and \emph{gluon} diagrams will eventually dominate. However, their contribution is suppressed by a factor $10$ due to a cancellation between the \emph{gluon} and \emph{twist-2} diagrams. When the scale of the LSP is such that coannihilation does not help anymore to maintain the correct relic density, the mixing is 
\be
N_{11}\simeq  \frac{(\sin\beta\pm\cos\beta)\sin\theta_\mathrm{w} M_Z}{\mp M_1 \sqrt{2}}\qquad N_{13}\simeq\pm\frac{1}{\sqrt{2}}\qquad N_{14}\simeq \frac{1}{\sqrt{2}},
\label{eq:purehiggsmallmix}
\ee
the \emph{Higgs} diagrams are suppressed by $M_1$ and the cross section decreases. Figure \ref{fig:UG} shows also the indirect reach from gluino searches: a 16 TeV gluino corresponds to $M_1\simeq 2.4$ TeV. This is the reach for a $100$ TeV collider in the $(m_{LSP},\sigma^{SI})$ plane. However, direct detection is stronger for $\mu>0$: indeed LZ can reach $M_1\simeq7$ TeV corresponding to gluinos around $42$ TeV. On the other hand the curve for negative $\mu$ is not bounded by LUX and LZ will explore pure higgsino states with $M_1\lesssim 1.7$ TeV. The gluino reach for a 100 TeV collider is stronger in this scenario. Continuing along the yellow curve, there is a value of $M_1$ such that the \emph{Higgs} contribution is of the same order of the \emph{gluon} and \emph{twist-2} diagrams and the cross section vanishes. Due to the cancellation uncertainty becomes $\mathcal{O}(1)$ and we cannot tell where exactly the cancellation happens. While the Bino completely decouples ($m_{\tilde{B}} =\mathcal{O}(100\,\,\mathrm{TeV})$), the \emph{Higgs} amplitude vanishes and the cross section reaches the value of the pure higgsino case given by the \emph{gluon}+\emph{twist-2} diagrams: 
\be
\sigma^{SI}_N \lesssim 10^{-48}\,\,\mathrm{cm}^2.	
\ee

When the LSP does not contribute to the whole DM abundance, the interplay between collider and direct dark matter searches is better shown in figure \ref{fig:muvsM1}. The dark blue area describes the region where the relic abundance exceeds the experimental value. 
The dark magenta is the bound by LUX, the light magenta region shows the reach of LZ and the light blue is the neutrino background. In red is the bound on charginos from LEP \cite{LEPSUSY}. In the plot are shown also future reach from mono-jet searches ($m_{\chi^0_1}\simeq 870$ GeV, \cite{Low:2014cba}) and the indirect reach coming from gluinos at a 100 TeV collider ($m_{\tilde{g}}\simeq 16$ TeV). 
According to the left panel of figure \ref{fig:muvsM1}, for $\mu<0$ thermal Bino-higgsino DM is not constrained by Direct Detection searches. Future experiments can however explore scenarios where $\mu\simeq -1.1$ TeV and $M_1< 1.7$ TeV. Nonetheless the strongest reach in this kind of models would come from a 100 TeV collider: gluino pair searches have the potential to explore a large area of the parameter space, while mono-jet searches will not have enough sensitivity to explore pure thermal higgsinos. The right panel shows how the Direct Detection reach is stronger for $\mu>0$. 

\begin{figure} [t]
\centering
\textbf{Split SUSY with universal gaugino masses}
\begin{center} 
\includegraphics[scale=0.59] {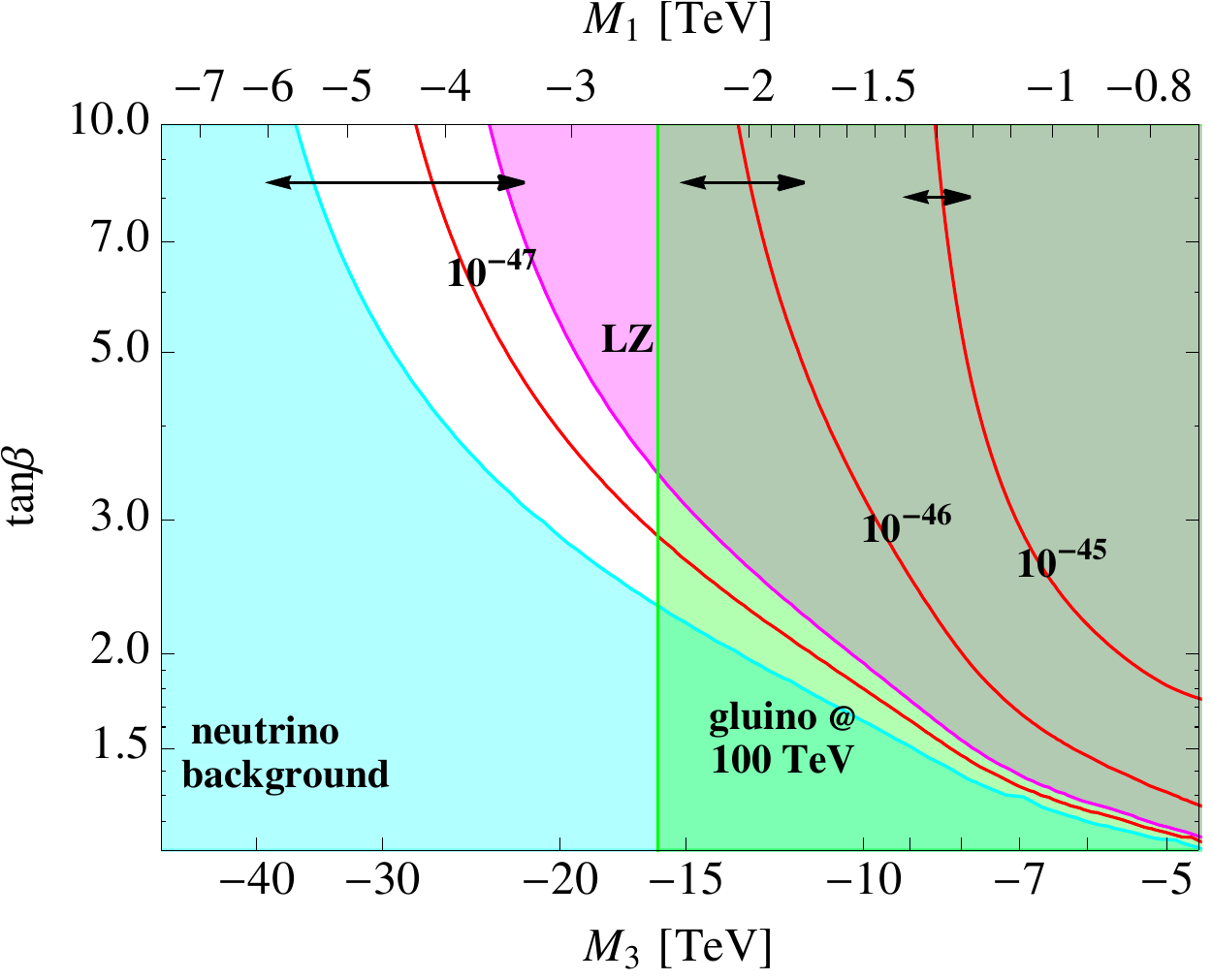}
\includegraphics[scale=0.59] {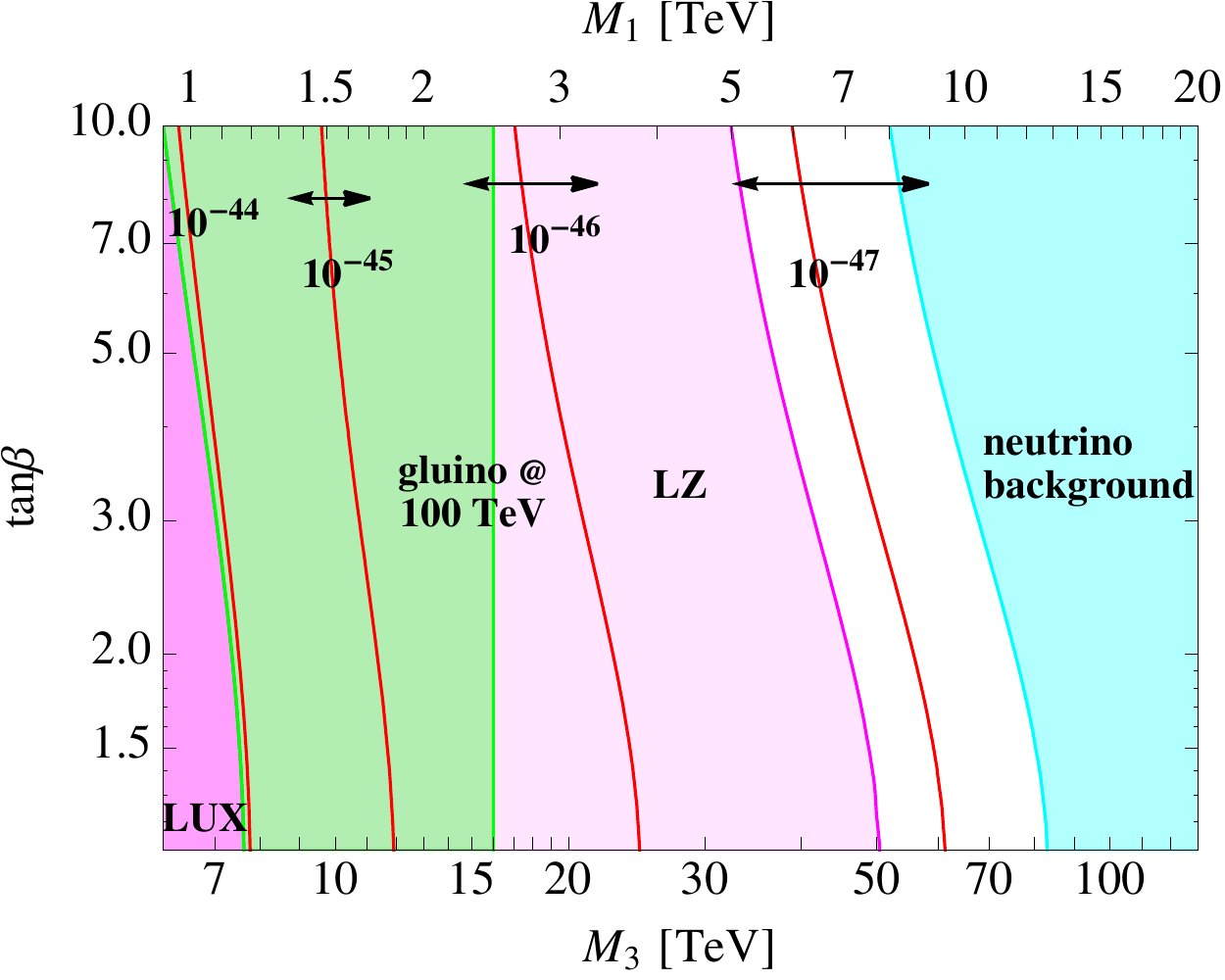}
\end{center}
\caption{Contours of SI neutralino-nucleon cross section in $[\mathrm{cm}^2]$ for higgsino dark matter (red curves). The black arrows shows the uncertainty on the cross section. The parameter $\mu$ is fixed such that the lightest neutralino has the correct relic density. The colour shading for actual bounds, future reach and background for direct detection are the same as in the previous figure. The green shaded area shows the expected reach at a $100$ TeV collider for gluino searches.} 
\label{fig:tanbvsM3} 
\end{figure}

The pure higgsino region is shown in figure \ref{fig:tanbvsM3} and it is interesting because it does not require the coincidence $|\mu-M_1|\ll|\mu|$ in order to explain the WIMP miracle. Figure \ref{fig:tanbvsM3} shows the dependence of $\tan\beta$ as a function of the gluino mass (and thus $M_1=\alpha_1/\alpha_3 M_3$), for a dark matter particle with the correct relic density for $M_3<0$ and $M_3>0$ in the left and right panel respectively (now we fixed $\mu$ to be positive and the sign of $M_3$ can vary). The value of the parameter $\mu$ has been fixed by requiring the correct relic density and it is approximately $1.1$ TeV across the whole plot. The colour labelling is the same as in the previous figures except for the green region that denotes the reach for gluinos at a 100 TeV collider. The red curves represent the SI cross section, while the black arrows show the uncertainties on the cross section. We notice that the collider reach is weaker than the direct detection experiments for $M_3>0$, while in the other case it can be competitive.


\subsection{Anomaly Mediation}
\label{sec:AMSBmodel}

In split SUSY with Anomaly Mediation \cite{Randall:1998uk,Giudice:1998xp,ArkaniHamed:2004yi} the physical gaugino masses are predicted in terms of the gravitino mass. The leading contributions to Bino and Wino masses come from one-loop anomaly mediation and threshold effects
\be
M_{1,2}=\frac{\beta_{1,2}}{g_{1,2}}m_{3/2}+\frac{\alpha_{1,2}}{2 \pi}   \frac{(\tilde{m}^2+\mu^2)\mu\tan\beta}{(\tan^2\beta+1)\tilde{m}^2+\mu^2}\ln\left[ (1+\tan^{-2}\beta) \left(  1+\frac{\tilde{m}^2}{\mu^2}  \right)  \right],
\label{eq:AM}
\ee 
where $g_i$ is the corresponding gauge coupling, $\beta_i$ its beta function, $m_{3/2}$ is the gravitino mass and $\tilde{m}$ is the scalar mass-scale. The gluino mass receives contributions only from anomaly mediation. In this scenario the scalars (except the SM-like Higgs) are heavy and close to the gravitino mass, while the gauginos are light. Higgsinos are not constrained. 

\begin{figure} [t]
\begin{center} 
\includegraphics[scale=0.59] {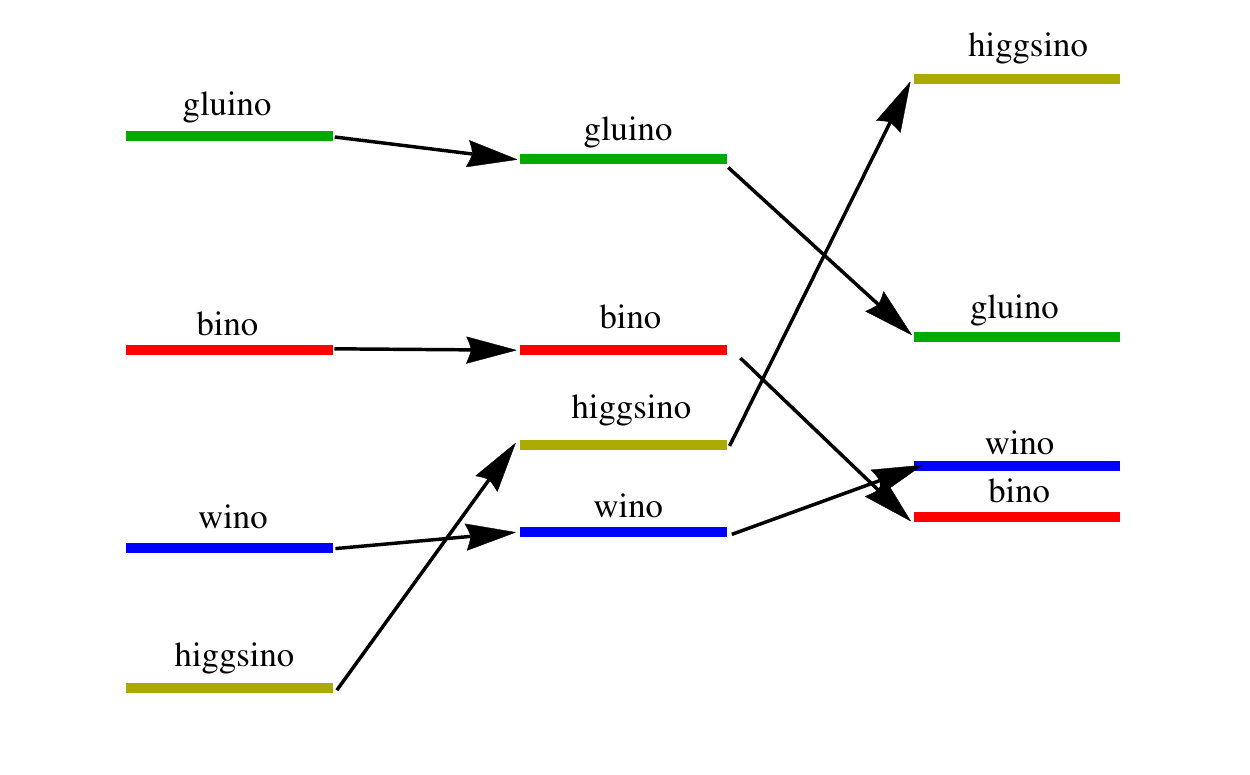}
\end{center}
\caption{Spectrum of gauginos and higgsinos in Anomaly Mediation.} 
\label{fig:spectrumAM} 
\end{figure}

Figure \ref{fig:spectrumAM} shows the spectrum of split SUSY with anomaly mediation. Depending on the contribution of the higgsinos the nature of the LSP changes. Light higgsinos lead to a spectrum in which the higgsino is the LSP and the ratio between gauginos is $M_1:M_2:M_3\simeq3:1:9$. Scenarios in which the Wino is the LSP are allowed if the higgsino is heavier than the Wino. For very heavy higgsinos the threshold corrections in \eqref{eq:AM} dominate and the Bino become the LSP.

 The model is entirely described in terms of four parameters: $m_{3/2}$, $\mu$, $\tan\beta$ and $\tilde{m}$. However, the value of the Higgs mass gives a relation between $\tan\beta$ and $\tilde{m}$. In order to have heavy scalars compatible with the Higgs mass we choose $\tan\beta=2$. We also discuss how the results change in the large $\tan\beta$ scenario.

\begin{figure} [t]
\centering%
\begin{minipage}{0.69\linewidth}
     \includegraphics[scale=0.85] {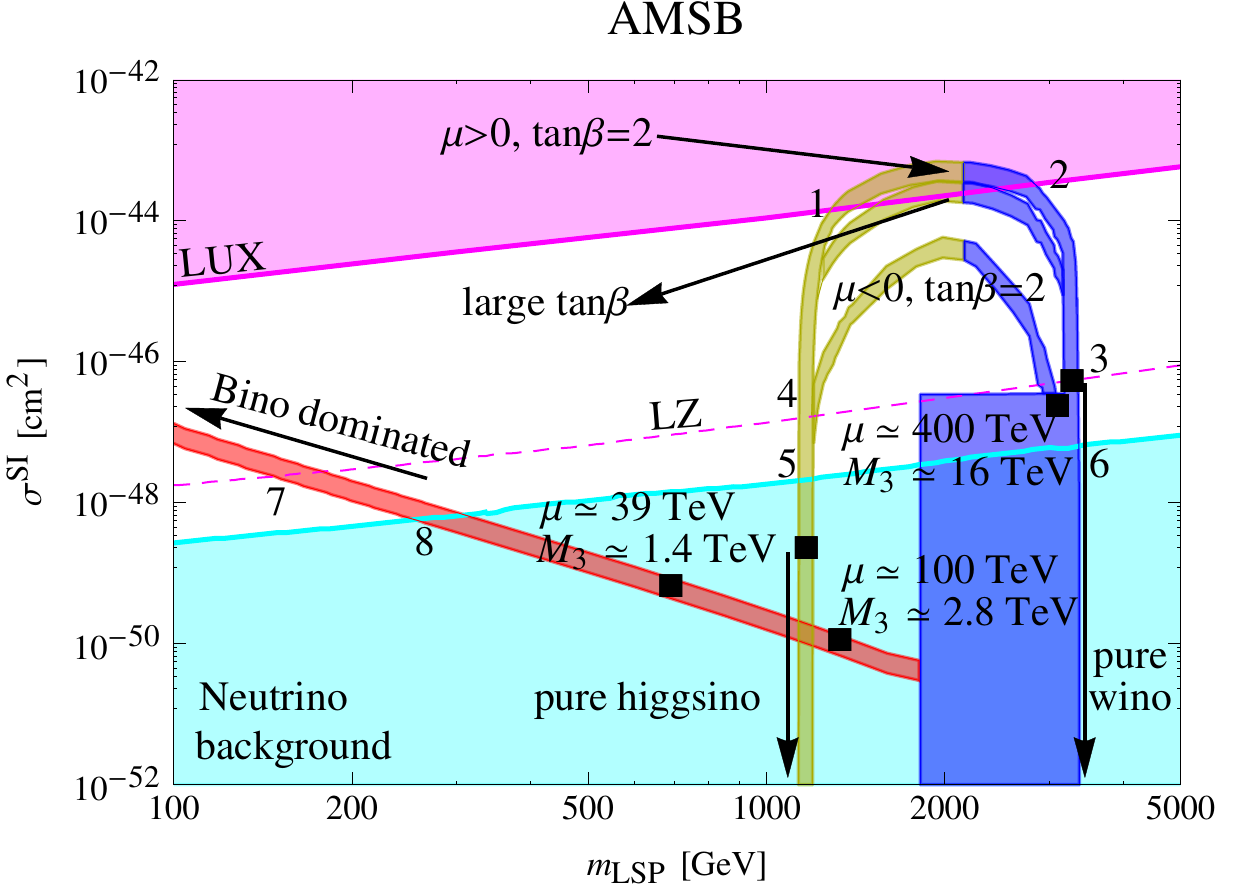}
   \end{minipage}
   \begin {minipage}{0.29\linewidth}
  \includegraphics[scale=0.5] {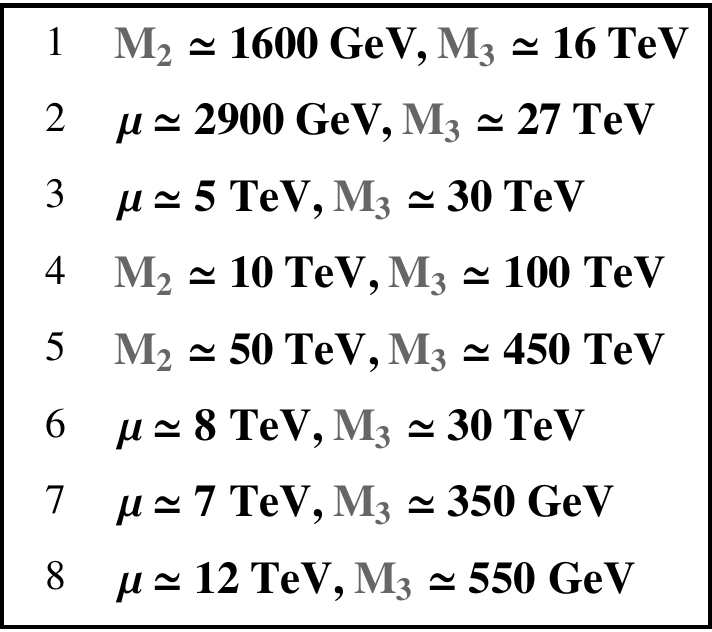}
   \end{minipage}
\caption{Spin independent nucleon-neutralino cross section in a split SUSY with anomaly mediation model (the red curve describe a Bino-like LSP, the yellow curve a higgsino-like and the blue one a Wino-like dark matter candidate). The magenta area is the actual bound by LUX, the dashed magenta line is the projected reach of LZ. The neutrino background is shaded in light blue. The numbers are explained in the legend on the right panel.} 
\label{fig:AMSB}
\end{figure}

\begin{figure} [t]
\centering
\textbf{AMSB}
\begin{center} 
\includegraphics[scale=0.59] {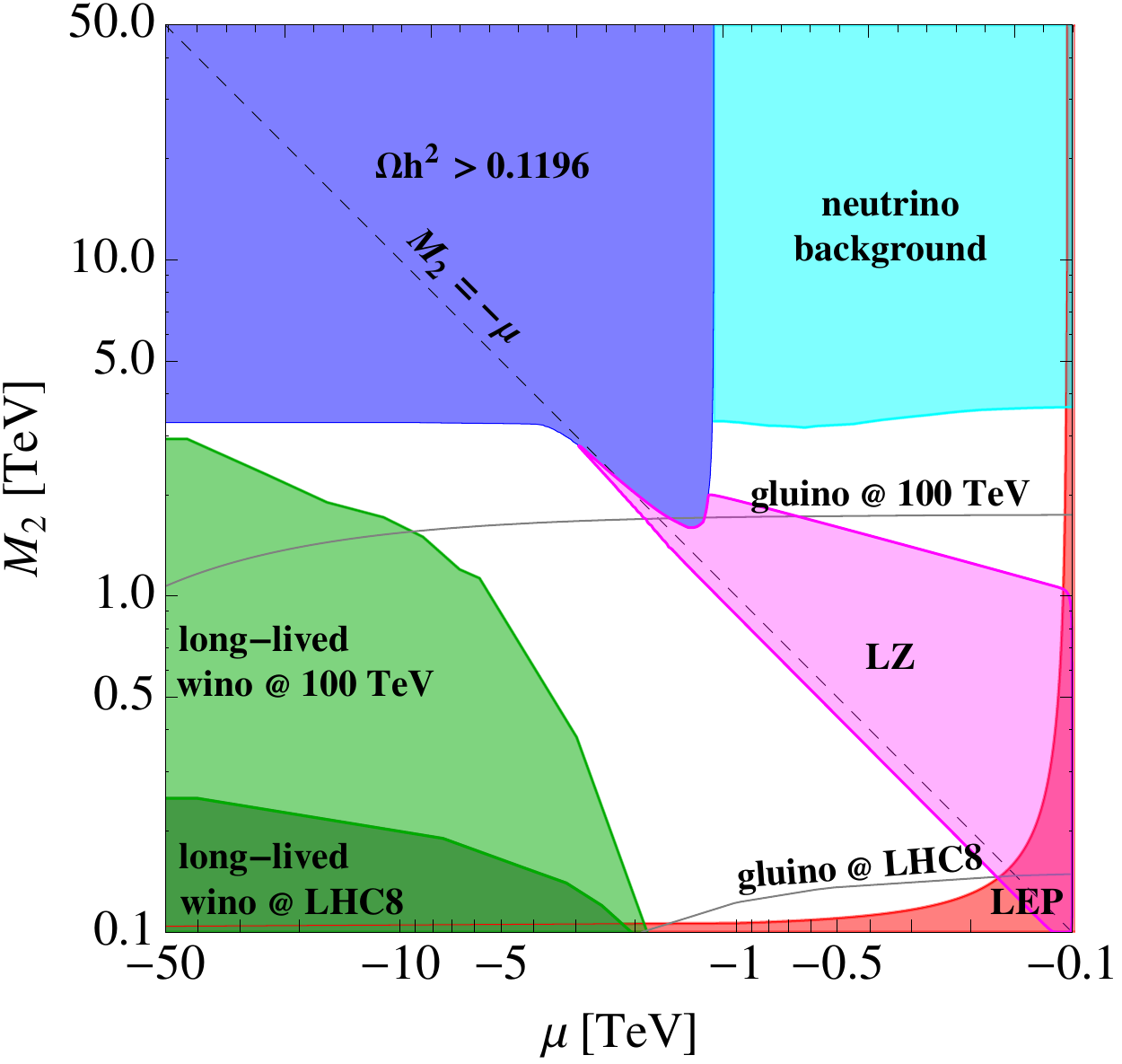}
\includegraphics[scale=0.59] {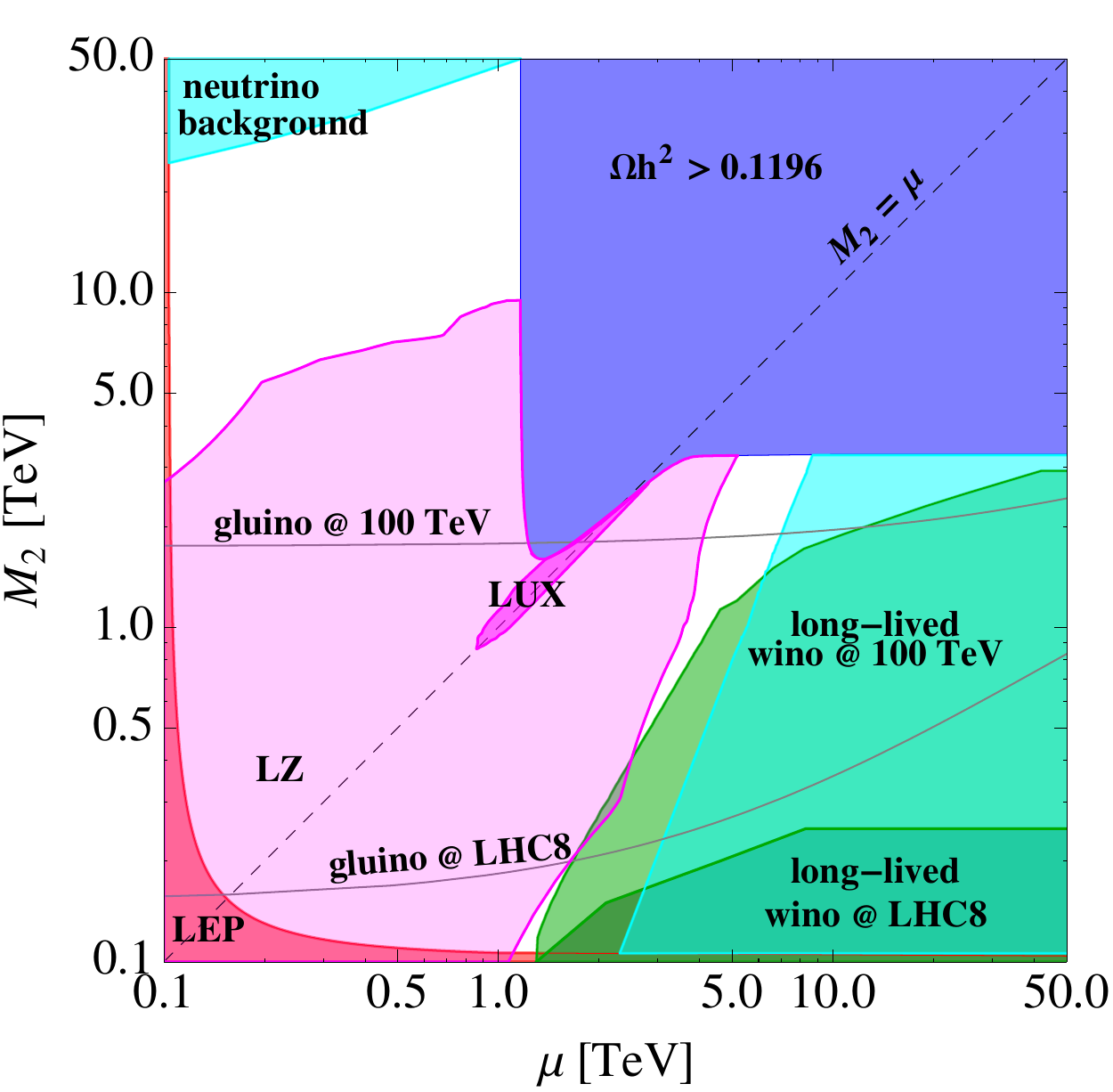}
\includegraphics[scale=0.59] {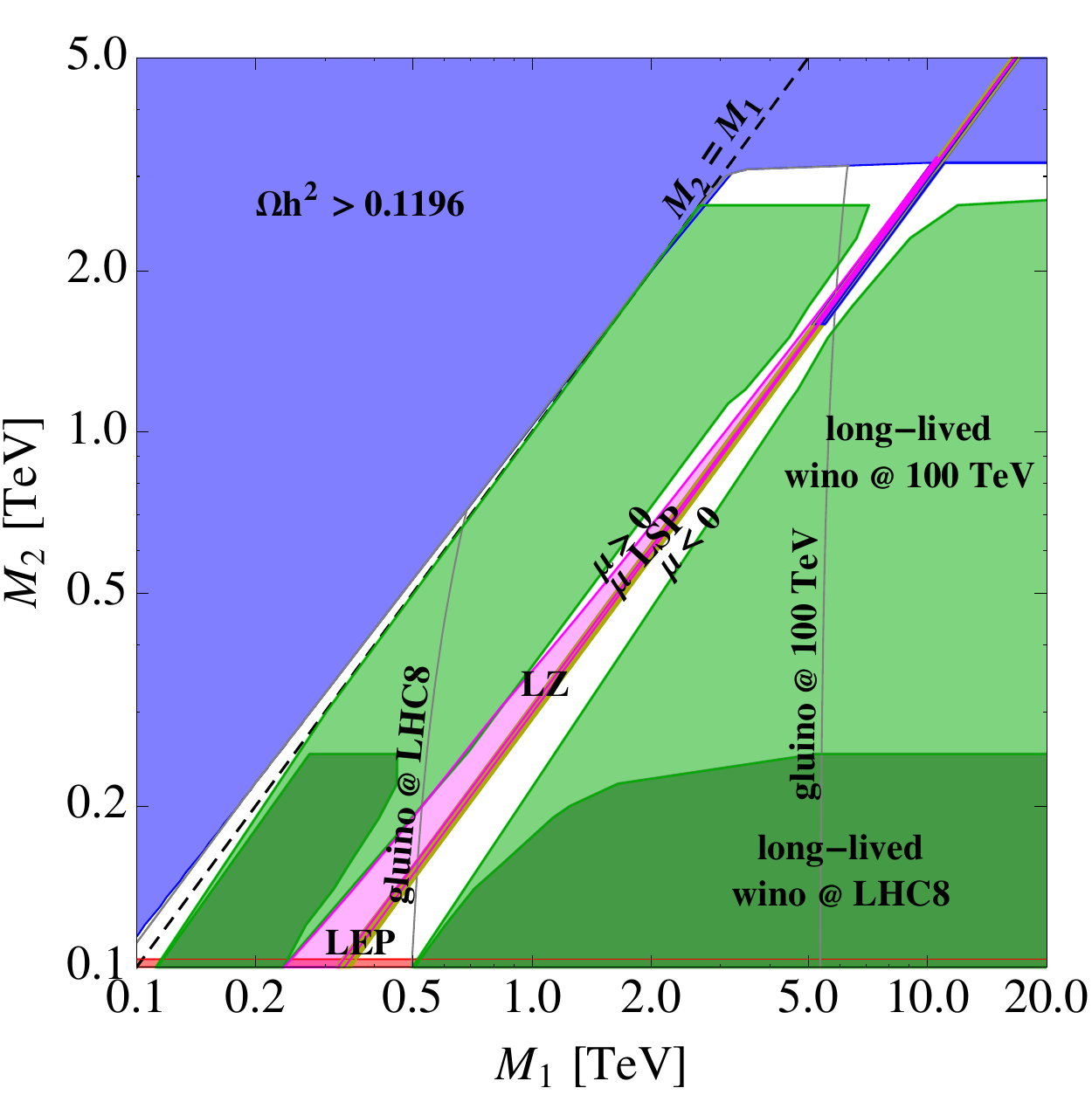}
\end{center}
\caption{The parameter space region allowed by the requirement that the neutralino relic abundance does not exceed the relic density in the plane  ($\mu$, $M_2$) for $\mu<0$ ($\mu>0$ ) in the left (right) panel. The bottom panel shows the plane ($M_1$, $M_2$). Direct Detection and collider constraints and future reach are also shown.} 
\label{fig:relicAMSB} 
\end{figure}

In figure \ref{fig:AMSB} we show the spin independent cross section that satisfies the relic density constraint. We start with the case $\mu<M_2$ and discuss the behaviour of the cross section and the bounds as the higgsino mass is increased. The yellow curve describes a mostly higgsino LSP state. At $m_{LSP}\simeq1.1$ TeV the cross section behaviour is the same explained for the Universal Gaugino masses scenario. The neutrino background makes it difficult for future direct detection experiments to probe this region of the parameter space, while LZ will probe only anomaly mediated spectra with higgsino LSP and $M_2 \lesssim 10$ TeV. As $M_2$ approaches $\mu$ the coupling increases and so does the cross section. A $100$ TeV collider may explore a very small region where $1.2\lesssim|\mu|\lesssim1.7$ TeV (see for example the gluino reaches of the top panels of figure \ref{fig:relicAMSB}). Continuing along the yellow curve of figure \ref{fig:AMSB}, when $\mu \sim M_2$ the mixing is maximal, the tree level Higgs exchange dominate and the LUX bounds apply. In this region the cross section is constant and the \emph{Higgs}  coupling is proportional to 
\be
(\cos\beta + \mathrm{sign}(\mu) \,\sin\beta).
\ee 
The lower curve represents the cross section for $\mu<0$ which is suppressed by the sign of $\mu$. At large $\tan\beta$ the sign of $\mu$ becomes irrelevant.
 Analogously to what happens for the pure higgsino case, the cross section decreases when $\mu\gg M_2$ and the LSP approaches the pure Wino state. Indeed now the mixing is given by $N_{12}\sim1$, 
 \be
 N_{13}\simeq \frac{\cos\theta_{\mathrm{w}}M_Z}{|\mu|^2}(M_2\cos\beta+\mu\sin\beta)\qquad  N_{14}\simeq \frac{\cos\theta_{\mathrm{w}}M_Z}{|\mu|^2}(M_2\sin\beta+\mu\cos\beta),
 \label{eq:smallmixwh/wb}
 \ee
and the Higgs coupling becomes
\be
-\frac{\cos\theta_{\mathrm{w}} M_Z}{|\mu|^2}(M_2+\mu\sin2\beta).
\label{eq:Wino}
\ee

The cross section therefore decreases as the higgsino decouples from the Wino. In this regime the \emph{gluon} and the \emph{twist-2} diagrams are also  important. However, as for the higgsino case the \emph{gluon} and the \emph{twist-2} amplitudes accidentally cancel suppressing their contribution by a factor $5$. Going down along the blue curve there is a value of $\mu$ for which the cross section vanish because the \emph{Higgs} diagrams cancel the \emph{gluon}$+$\emph{twist-2} contributions. Due to $\mathcal{O}(1)$ uncertainties it is not possible to define exactly for which value of $\mu$ this cancellation happens. When the higgsino and the Bino are both decoupled, the cross section is estimated to be
\be
\sigma_{N}^{SI}\lesssim 10^{-47}\,\,\mathrm{cm}^2.
\ee 
If one keeps increasing the value of $\mu$, with $M_2\sim3$ TeV in order to reproduce the correct relic density that provide a Wino dark matter candidate, the mass of the Bino-like neutralino decreases (see \eqref{eq:AM}). While the splitting between Wino and Bino decreases, the cross section increases because the \emph{Higgs} diagrams become negligible with respect to the other contributions ($N_{11}$ is negligible with respect to $N_{12}$ and $N_{13}$ and $N_{14}$ are given by equation \eqref{eq:smallmixwh/wb}). This is the top flat edge of the blue rectangle. In this region the LSP neutralino is a pure Wino, with $M_1$ closer and closer and $\mu$ decoupled. From the top right to the top left part of the blue rectangle, the value of $M_1$ ranges between $10$ and $1.8$ TeV, while the higgsino is decoupled and the LSP mass is given by the value of $M_2$. The mixing between Bino and Wino is always negligible for mass splitting larger than a GeV. Once $M_1<M_2$ the \emph{Higgs} diagrams, the only contributions to the cross section, become suppressed by the large value of $\mu$. In this region the neutralino is almost a pure Bino with the mixing given by \eqref{eq:purebinomix}. In this case in order to have the correct relic, the Bino must coannihilate with the Wino and therefore the splitting must be $\lesssim 30$ GeV. In order to decrease the gaugino mass scale and maintain such splitting, $\mu$ has to decrease and thus the cross section increases, being the Higgs coupling given by
\be
M_W\tan\theta_{\mathrm{w}} \frac{M_1+\mu\sin2\beta}{|\mu|^2}.
\ee 
There is no top threshold in this case because the annihilation into $t\bar{t}$ is not the dominant contribution. A $100$ TeV collider could be able to explore the whole region where the LSP is a mixed Bino/higgsino and Bino/Wino state from the LEP bound to $M_{LSP} \sim  3.1$ TeV (those reaches are better shown in the following figure \ref{fig:relicAMSB}). Given the large value of $\mu$, at large $\tan\beta$ the cross section is further suppressed.

The ($\mu$, $M_2$) and the ($M_1$, $M_2$) planes are shown in figure \ref{fig:relicAMSB}. 
The upper panels show the ($\mu$, $M_2$) planes for $\mu<0$ (left) and $\mu>0$ (right). The blue region is excluded by the requirement on the relic density. The dark and light green areas describe the constraints from long lived Winos at LHC8 and the future reach for a 100 TeV collider. The grey lines shows the bounds and reach on gluino pair searches. It is interesting to note that the direct detection reach is limited in the left panel, due to the suppression of the cross section for $\mu<0$. On the right panel the two different searches are complementary. The bottom panel shows the plane ($M_1$, $M_2$), for both positive and negative $\mu$. The colour coding is the same as in the upper plots. The yellow line that cuts the panel in two represents the area in which the higgsino is the LSP and it cuts the plane in a region with $\mu>0$ (left) and $\mu<0$ (right). The yellow region is strongly connected with the upper plots. The blue stripe overlapping the yellow line is a region in which $M_1$ is decoupled and it shows the crossing between the Wino LSP parameter space and the higgsino one. A large region of the parameter space could be probed at a 100 TeV collider, leaving unexplored just the narrow region corresponding to the pure Wino and higgsino cases. 


\section{Conclusions}

In this paper, we have presented the mass reach of several electroweakino searches for future hadron colliders, their implications for DM and the complementarity with direct detection experiments. 

In particular we have studied scenarios where Wino NLSPs decay into leptons (or $b$-jets and leptons) and Bino LSP, with long-lived charged Winos in models of anomaly mediation and with Wino or higgsino LSP in GMSB models. Concerning direct detection experiments we analysed split SUSY with universal gaugino masses and models of anomaly mediation. We analysed both the cases in which the lightest neutralino contributes entirely or partially to the dark matter abundance.

The LHC excludes electroweakinos up to few hundred GeV, well below the interesting cases of pure Wino or higgsino dark matter. In addition, current bounds from LUX are stronger in models with universal gaugino masses and positive $\mu$ but are non-existent or very weak in the other scenarios studied. 

Electroweakino collider searches are relevant, for example, in low scale gauge mediation models with universal gaugino masses and large $\tan\beta$ or in models of anomaly mediation. Indeed in the first scenario the gluino is expected to be at the same scale or below the squarks (the Higgs mass fixes this scale to be around $10$ TeV) that means a Wino around $3$ TeV. Such Winos could be explored at a 100 TeV collider with less than 100 fb$^{-1}$ of luminosity. In anomaly mediation the ratio between the gluino and the Wino mass is large, making the Wino searches more powerful: a 100 TeV collider with few ab$^{-1}$ of luminosity could explore $3$ TeV Winos ($m_{\mathrm{gluino}}\gtrsim 20$ TeV).

In split SUSY models with universal gaugino masses the strongest mass reach comes from direct detection in the positive $\mu$ scenario. The combination of direct detection experiments, and monojet and gluino searches at a 100 TeV collider will leave unexplored a narrow region with $\mu\gtrsim870$ GeV and $M_1\gtrsim5$ TeV. On the other hand in the negative $\mu$ case direct detection experiments are weaker, due to the suppression of the cross section. However, gluino and monojet searches may explore also the region where the neutrino background limits direct detection experiments. In this scenario, therefore, it would be possible to explore the entire parameter space where the (non-termal) neutralino dark matter candidate is an admixture of higgsino and Bino states.

In anomaly mediation models direct detection experiments and collider searches are complementary. Indeed searches on long-lived Winos set strong mass reach in regions where direct detection is weak. The interplay between gluino searches, long-lived Winos and direct detection may cover large area of the parameter space where the neutralino does not contribute to the whole dark matter. Gluino searches at a 100 TeV collider could be able to explore the whole parameter space in which the dark matter particle is a Bino/higgsino or Bino/Wino mixed state. Only in the case with positive $\mu$, LZ may probe pure thermal Winos or higgsinos.

In conclusion we showed what are the prospects for neutralino dark matter in future direct detection experiments and collider searches. We also showed that direct bounds from electroweakino searches are not always stronger than the bounds from gluino pair production. Moreover we showed the interplay and complementarity between the two kind of searches, indicating that both have great potential for discovering dark matter. As a final remark it is interesting to note that at the LHC13 at least $10$ fb$^{-1}$ of luminosity (expected by the end of 2015) are needed in order to match the current limits from the LHC8 on the electroweakino searches discussed. Moreover in the low mass region an increase of the luminosity of a factor ten might have the same effect of increasing the centre of mass energy of a factor two or more.

We are still far from exploring all the parameter space for Wino and higgsino dark matter but a 100 TeV collider seems to be a necessary tool in order to achieve this goal.

\acknowledgments
We would like to thank G.Villadoro and P.Ullio for helpful discussions and G.Villadoro for carefully reading of the manuscript.  We also thank A.Hryczuk for providing a prerelease version 1.1 of his DarkSE code. We acknowledge ICTP for hospitality and for allowing the use of the Argo cluster and I.Girotto for providing technical support.
 
\paragraph{Note added:} after the completion of this work Ref. \cite{Bramante:2014tba} appeared with partial overlap with our results.


\begin{thebibliography}{99}

\bibitem{Aad:2012tfa}
  G.~Aad {\it et al.}  [ATLAS Collaboration],
  \emph{Observation of a new particle in the search for the Standard Model Higgs boson with the ATLAS detector at the LHC},
  Phys.\ Lett.\ B {\bf 716} (2012) 1
    [\href{http://xxx.lanl.gov/abs/1207.7214}{{\tt arXiv:1207.7214}}].

\bibitem{Chatrchyan:2012ufa}
  S.~Chatrchyan {\it et al.}  [CMS Collaboration],
  \emph{Observation of a new boson at a mass of 125 GeV with the CMS experiment at the LHC},
  Phys.\ Lett.\ B {\bf 716} (2012) 30
    [\href{http://xxx.lanl.gov/abs/1207.7235}{{\tt arXiv:1207.7235}}].

\bibitem{Craig:2013cxa}
  N.~Craig,
 \emph{The State of Supersymmetry after Run I of the LHC},
    [\href{http://arxiv.org/abs/1309.0528v2}{{\tt  arXiv:1309.0528 [hep-ph]}}].

\bibitem{Wells:2003tf}
  J.~D.~Wells,
  \emph{Implications of supersymmetry breaking with a little hierarchy between gauginos and scalars},
   [\href{http://arxiv.org/pdf/hep-ph/0306127.pdf}{{\tt hep-ph/0306127}}].

\bibitem{ArkaniHamed:2004fb}
  N.~Arkani-Hamed and S.~Dimopoulos,
  \emph{Supersymmetric unification without low energy supersymmetry and signatures for fine-tuning at the LHC},
  JHEP {\bf 0506} (2005) 073
    [\href{http://xxx.lanl.gov/abs/hep-th/0405159}{{\tt hep-th/0405159}}].

\bibitem{Giudice:2004tc}
  G.~F.~Giudice and A.~Romanino,
  \emph{Split supersymmetry},
  Nucl.\ Phys.\ B {\bf 699} (2004) 65
   [Erratum-ibid.\ B {\bf 706} (2005) 65]
    [\href{http://xxx.lanl.gov/abs/hep-ph/0406088}{{\tt hep-ph/0406088}}].

\bibitem{ArkaniHamed:2004yi}
  N.~Arkani-Hamed, S.~Dimopoulos, G.~F.~Giudice and A.~Romanino,
  \emph{Aspects of split SUSY},
  Nucl.\ Phys.\ B {\bf 709} (2005) 3
  [\href{http://xxx.lanl.gov/abs/hep-ph/0409232}{{\tt hep-ph/0409232}}].

\bibitem{Pierce:2004mk}
  A.~Pierce,
  \emph{Dark matter in the finely tuned minimal supersymmetric standard model},
  Phys.\ Rev.\ D {\bf 70} (2004) 075006
  [\href{http://xxx.lanl.gov/abs/hep-ph/0406144}{{\tt hep-ph/0406144}}].
  
  \bibitem{Masiero:2004ft}
  A.~Masiero, S.~Profumo and P.~Ullio,
  \emph{Neutralino dark matter detection in split supersymmetry scenarios},
  Nucl.\ Phys.\ B {\bf 712} (2005) 86
    [\href{http://xxx.lanl.gov/abs/hep-ph/0412058}{{\tt hep-ph/0412058}}].
  
  \bibitem{Baer:2005zc}
  H.~Baer, A.~Mustafayev, E.~K.~Park and S.~Profumo,
  \emph{Mixed wino dark matter: Consequences for direct, indirect and collider detection},
  JHEP {\bf 0507} (2005) 046
  [\href{http://xxx.lanl.gov/abs/hep-ph/0505227}{{\tt hep-ph/0505227}}].

\bibitem{Baer:2005jq}
  H.~Baer, T.~Krupovnickas, A.~Mustafayev, E.~K.~Park, S.~Profumo and X.~Tata,
  \emph{Exploring the BWCA (bino-wino co-annihilation) scenario for neutralino dark matter},
  JHEP {\bf 0512} (2005) 011
   [\href{http://xxx.lanl.gov/abs/hep-ph/0511034}{{\tt hep-ph/0511034}}].

\bibitem{BirkedalHansen:2001is}
  A.~Birkedal-Hansen and B.~D.~Nelson,
  \emph{The Role of Wino content in neutralino dark matter},
  Phys.\ Rev.\ D {\bf 64} (2001) 015008
  [\href{http://xxx.lanl.gov/abs/hep-ph/0102075}{{\tt hep-ph/0102075}}].

\bibitem{BirkedalHansen:2002am}
  A.~Birkedal-Hansen {\it et al.}  [Michigan Center for Theoretical Physics, U. of Michigan Collaboration],
  \emph{Relic neutralino densities and detection rates with nonuniversal gaugino masses},
  Phys.\ Rev.\ D {\bf 67} (2003) 095006
  [\href{http://xxx.lanl.gov/abs/hep-ph/0211071}{{\tt hep-ph/0211071}}].

\bibitem{ArkaniHamed:2006mb}
  N.~Arkani-Hamed, A.~Delgado and G.~F.~Giudice,
\emph{The Well-tempered neutralino},
  Nucl.\ Phys.\ B {\bf 741} (2006) 108
    [\href{http://xxx.lanl.gov/abs/hep-ph/0601041}{{\tt hep-ph/0601041}}].
   
\bibitem{Aad:2014nua}
  G.~Aad {\it et al.}  [ATLAS Collaboration],
  \emph{Search for direct production of charginos and neutralinos in events with three leptons and missing transverse momentum in $\sqrt{s} =$ 8TeV $pp$ collisions with the ATLAS detector},
  \emph{JHEP} {\bf 1404} (2014) 169
 [\href{http://xxx.lanl.gov/abs/1402.7029}{{\tt arXiv:1402.7029}}].

\bibitem{Khachatryan:2014qwa}
  V.~Khachatryan {\it et al.}  [CMS Collaboration],
  \emph{Searches for electroweak production of charginos, neutralinos, and sleptons decaying to leptons and W, Z, and Higgs bosons in pp collisions at 8 TeV},
  \emph{Eur.\ Phys.\ J.\ C} {\bf 74} (2014) 9,  3036
   [\href{http://xxx.lanl.gov/abs/1405.7570}{{\tt arXiv:1405.7570}}].
  
  \bibitem{Calibbi:2013poa}
  L.~Calibbi, J.~M.~Lindert, T.~Ota and Y.~Takanishi,
  \emph{Cornering light Neutralino Dark Matter at the LHC},
  JHEP {\bf 1310} (2013) 132
  [\href{http://xxx.lanl.gov/abs/1307.4119}{{\tt arXiv:1307.4119}}].
  
  \bibitem{Calibbi:2014lga}
  L.~Calibbi, J.~M.~Lindert, T.~Ota and Y.~Takanishi,
  \emph{LHC Tests of Light Neutralino Dark Matter without Light Sfermions},
  JHEP {\bf 1411} (2014) 106
  [\href{http://xxx.lanl.gov/abs/1410.5730}{{\tt arXiv:1410.5730}}].
  
  \bibitem{ATLAS-PUB-2014-010}
  The ATLAS Collaboration,
  \emph{Search for SUSY at the high luminosity LHC with the ATLAS Detector}, Tech. Rep.
   ATLAS-PHYS-PUB-2014-010, CERN, Geneva, 2014.
  
  \bibitem{CMS-PAS-FTR-13-2014}
  The CMS Collaboration
  \emph{Study of the Discovery Reach in Searches for Supersymmetry at CMS with $3000$ fb$^{-1}$}, Tech. Rep.
  CMS-PAS-FTR-13-2014, CERN, Geneva, 2013

\bibitem{Xiao:2014xyn}
PandaX Collaboration, M.~Xiao et~al., {\it {First dark matter search
  results from the PandaX-I experiment}},  {\em Sci.China Phys.Mech.Astron.}
  {\bf 57} (2014) 2024--2030, [\href{http://arxiv.org/abs/1408.5114}{{\tt
  arXiv:1408.5114}}].

\bibitem{Agnese:2014aze}
 SuperCDMS Collaboration, R.~Agnese et~al., {\it {Search for Low-Mass
  WIMPs with SuperCDMS}},  {\em Phys.Rev.Lett.} {\bf 112} (2014) 241302,
  [\href{http://arxiv.org/abs/1402.7137}{{\tt arXiv:1402.7137}}].

\bibitem{Akerib:2013tjd}
 LUX Collaboration, D.~Akerib et~al., {\it {First results from the LUX
  dark matter experiment at the Sanford Underground Research Facility}},  {\em
  Phys.Rev.Lett.} {\bf 112} (2014) 091303,
  [\href{http://arxiv.org/abs/1310.8214}{{\tt arXiv:1310.8214}}].

\bibitem{Aprile:2012nq}
 XENON100 Collaboration, E.~Aprile et~al., {\it {Dark Matter Results from
  225 Live Days of XENON100 Data}},  {\em Phys.Rev.Lett.} {\bf 109} (2012)
  181301, [\href{http://arxiv.org/abs/1207.5988}{{\tt arXiv:1207.5988}}].

\bibitem{Aprile:2013doa}
XENON100 Collaboration, E.~Aprile et~al., {\it {Limits on spin-dependent
  WIMP-nucleon cross sections from 225 live days of XENON100 data}},  {\em
  Phys.Rev.Lett.} {\bf 111} (2013), no.~2 021301,
  [\href{http://arxiv.org/abs/1301.6620}{{\tt arXiv:1301.6620}}].

\bibitem{2011PhRvL.107x1302A}
FERMI--LAT Collaboration, M.~Ackermann et~al., {\it {Costraining Dark Matter Models from a Combined Analysis of Milky Way Satellites with the Fermi Area Telescope}},  {\em
  Phys.Rev.Lett.} {\bf 107} (2011), 241302,
  [\href{http://arxiv.org/abs/1108.3546}{{\tt arXiv:1108.3546}}].

\bibitem{2010PhRvL.105l1101A}
PAMELA Collaboration, O.~Adriani et~al., {\it {PAMELA results on the cosmic-ray antiproton flux from 60 MeV to 180 GeV in kinetic energy}},  {\em
  Phys.Rev.Lett.} {\bf 105} (2010), 121101,
  [\href{http://arxiv.org/abs/1007.0821}{{\tt arXiv:1007.0821}}].

\bibitem{Cohen:2013ama}
  T.~Cohen, M.~Lisanti, A.~Pierce and T.~R.~Slatyer,
  \emph{Wino Dark Matter Under Siege},
  JCAP {\bf 1310} (2013) 061
    [\href{http://xxx.lanl.gov/abs/1307.4082}{{\tt arXiv:1307.4082}}].

\bibitem{Fan:2013faa}
  J.~Fan and M.~Reece,
\emph{In Wino Veritas? Indirect Searches Shed Light on Neutralino Dark Matter},
  JHEP {\bf 1310} (2013) 124
    [\href{http://xxx.lanl.gov/abs/1307.4400}{{\tt arXiv:1307.4400}}].

\bibitem{Hryczuk:2014hpa}
  A.~Hryczuk, I.~Cholis, R.~Iengo, M.~Tavakoli and P.~Ullio,
  \emph{Indirect Detection Analysis: Wino Dark Matter Case Study},
  JCAP {\bf 1407} (2014) 031
    [\href{http://xxx.lanl.gov/abs/1401.6212}{{\tt arXiv:1401.6212}}].

\bibitem{Cohen:2013xda}
  T.~Cohen, T.~Golling, M.~Hance, A.~Henrichs, K.~Howe, J.~Loyal, S.~Padhi and J.~G.~Wacker,
  \emph{SUSY Simplified Models at 14, 33, and 100 TeV Proton Colliders},
  JHEP {\bf 1404} (2014) 117
    [\href{http://xxx.lanl.gov/abs/1311.6480}{{\tt arXiv:1311.6480}}].

\bibitem{Cohen:2014hxa}
  T.~Cohen, R.~T.~D'Agnolo, M.~Hance, H.~K.~Lou and J.~G.~Wacker,
  \emph{Boosting Stop Searches with a 100 TeV Proton Collider},
  JHEP {\bf 1411} (2014) 021
    [\href{http://xxx.lanl.gov/abs/1406.4512}{{\tt arXiv:1406.4512 [hep-ph]}}].

\bibitem{Zhou:2013raa}
  N.~Zhou, D.~Berge, L.~Wang, D.~Whiteson and T.~Tait,
  \emph{Sensitivity of future collider facilities to WIMP pair production via effective operators and light mediators},
    [\href{http://xxx.lanl.gov/abs/1307.5327}{{\tt arXiv:1307.5327}}].

\bibitem{Low:2014cba}
  M.~Low and L.~T.~Wang,
  \emph{Neutralino dark matter at 14 TeV and 100 TeV},
  JHEP {\bf 1408} (2014) 161
    [\href{http://xxx.lanl.gov/abs/1404.0682}{{\tt arXiv:1404.0682}}].

\bibitem{Cirelli:2014dsa}
  M.~Cirelli, F.~Sala and M.~Taoso,
  \emph{Wino-like Minimal Dark Matter and future colliders},
  JHEP {\bf 1410} (2014) 033
    [\href{http://xxx.lanl.gov/abs/1407.7058}{{\tt arXiv:1407.7058}}].

\bibitem{Acharya:2014pua}
  B.~S.~Acharya, K.~Bozek, C.~Pongkitivanichkul and K.~Sakurai,
  \emph{Prospects for observing charginos and neutralinos at a 100 TeV proton-proton collider},
        [\href{http://xxx.lanl.gov/abs/1410.1532}{{\tt arXiv:1410.1532}}].

\bibitem{Gori:2014oua}
  S.~Gori, S.~Jung, L.~T.~Wang and J.~D.~Wells,
  \emph{Prospects for Electroweakino Discovery at a 100 TeV Hadron Collider},
  [\href{http://xxx.lanl.gov/abs/1410.6287}{{\tt arXiv:1410.6287}}].

\bibitem{Baltz:2004aw}
  E.~A.~Baltz and P.~Gondolo,
  \emph{Markov chain Monte Carlo exploration of minimal supergravity with implications for dark matter},
  JHEP {\bf 0410} (2004) 052
    [\href{http://xxx.lanl.gov/abs/hep-ph/0407039}{{\tt hep-ph/0407039}}].

\bibitem{Cheung:2012qy}
  C.~Cheung, L.~J.~Hall, D.~Pinner and J.~T.~Ruderman,
  \emph{Prospects and Blind Spots for Neutralino Dark Matter},
  JHEP {\bf 1305} (2013) 100
    [\href{http://xxx.lanl.gov/abs/1211.4873}{{\tt arXiv:1211.4873}}].

\bibitem{Han:2013gba}
  T.~Han, Z.~Liu and A.~Natarajan,
  \emph{Dark matter and Higgs bosons in the MSSM},
  JHEP {\bf 1311} (2013) 008
    [\href{http://xxx.lanl.gov/abs/1303.3040}{{\tt arXiv:1303.3040}}].
  
  \bibitem{Cheung:2013dua}
  C.~Cheung and D.~Sanford,
  \emph{Simplified Models of Mixed Dark Matter},
  JCAP {\bf 1402} (2014) 011
    [\href{http://xxx.lanl.gov/abs/1311.5896}{{\tt arXiv:1311.5896}}].
  
  \bibitem{Huang:2014xua}
  P.~Huang and C.~E.~M.~Wagner,
  \emph{Blind Spots for neutralino Dark Matter in the MSSM with an intermediate $m_A$},
  Phys.\ Rev.\ D {\bf 90} (2014) 015018
    [\href{http://xxx.lanl.gov/abs/1404.0392}{{\tt arXiv:1404.0392}}].
    
\bibitem{Han:2014nba}
  T.~Han, Z.~Liu and S.~Su,
  \emph{Light Neutralino Dark Matter: Direct/Indirect Detection and Collider Searches},
  JHEP {\bf 1408} (2014) 093
    [\href{http://xxx.lanl.gov/abs/1406.1181}{{\tt arXiv:1406.1181}}].

\bibitem{ATLAS2013-093}
  G.~Aad {\it et al.}  [ATLAS Collaboration],
  \emph{Search for chargino and neutralino production in final states with one lepton, two b-jets consistent with a Higgs boson, and missing transverse momentum with the ATLAS detector in $20.3fb^{-1}$ of $\sqrt{s}=8TeV$ $pp$ collisions}, Tech. Rep. 
  ATLAS-CONF-2013-093, CERN, Geneva, 2014.

\bibitem{Aad:2013yna}
  G.~Aad {\it et al.}  [ATLAS Collaboration],
  \emph{Search for charginos nearly mass degenerate with the lightest neutralino based on a disappearing-track signature in pp collisions at  $\sqrt(s)$=8 TeV with the ATLAS detector},
 \emph{Phys.\ Rev.\ D}{\bf 88} (2013) 11,  112006
  [\href{http://xxx.lanl.gov/abs/1310.3675}{{\tt arXiv:1310.3675}}].

\bibitem{ATLAS-2014-001}
  The ATLAS collaboration,
  \emph{Search for Diphoton Events with Large Missing Transverse Momentum in 8 TeV pp Collision Data with the ATLAS Detector},
  ATLAS-CONF-2014-001

\bibitem{CR}
  G.~Salam, A.~Weiler,
  \href{http://collider-reach.web.cern.ch}{{\tt  http://collider-reach.web.cern.ch}}.

\bibitem{CR2}
  G.~Salam, A.~Weiler,
 \emph{Collider reach $\beta$},
       \href{https://gsalam.web.cern.ch/gsalam/talks/repo/2014-02-FHC-PU-collider-reach.pdf}{{\tt  https://gsalam.web.cern.ch/gsalam/talks/repo/2014-02-FHC-PU-collider-reach.pdf}}.

\bibitem{Debove:2008nr}
  J.~Debove, B.~Fuks and M.~Klasen,
  \emph{Model-independent analysis of gaugino-pair production in polarized and unpolarized hadron collisions},
  \emph{Phys.\ Rev.\ D} {\bf 78} (2008) 074020
   [\href{http://xxx.lanl.gov/abs/0804.0423}{{\tt arXiv:0804.0423}}].

\bibitem{Martin:2009iq}
  A.~D.~Martin, W.~J.~Stirling, R.~S.~Thorne and G.~Watt,
  \emph{Parton distributions for the LHC},
  Eur.\ Phys.\ J.\ C {\bf 63} (2009) 189
    [\href{http://xxx.lanl.gov/abs/0901.0002}{{\tt arXiv:0901.0002}}].



\bibitem{Cheng:1998hc}
  H.~C.~Cheng, B.~A.~Dobrescu and K.~T.~Matchev,
  \emph{Generic and chiral extensions of the supersymmetric standard model},
  Nucl.\ Phys.\ B {\bf 543} (1999) 47
  [\href{http://xxx.lanl.gov/abs/hep-ph/9811316}{{\tt hep-ph/9811316}}].

\bibitem{Feng:1999fu}
  J.~L.~Feng, T.~Moroi, L.~Randall, M.~Strassler and S.~f.~Su,
  \emph{Discovering SUSY at the Tevatron in wino LSP scenarios},
  Phys.\ Rev.\ Lett.\  {\bf 83} (1999) 1731
   [\href{http://xxx.lanl.gov/abs/hep-ph/9904250}{{\tt hep-ph/9904250}}].

\bibitem{Gherghetta:1999sw}
  T.~Gherghetta, G.~F.~Giudice and J.~D.~Wells,
  \emph{Phenomenological consequences of SUSY with anomaly induced masses},
  Nucl.\ Phys.\ B {\bf 559} (1999) 27
 [\href{http://xxx.lanl.gov/abs/hep-ph/9904378}{{\tt hep-ph/9904378}}] .


  
  \bibitem{Campbell:2003dd}
  J.~M.~Campbell, R.~K.~Ellis, F.~Maltoni and S.~Willenbrock,
  \emph{Associated production of a $Z$ Boson and a single heavy quark jet},
  Phys.\ Rev.\ D {\bf 69} (2004) 074021
  [\href{http://xxx.lanl.gov/abs/hep-ph/0312024}{{\tt hep-ph/0312024}}].

\bibitem{Haas}
  A.~Haas,
  \emph{LHC SUSY searches: Pulling out all the stops},
 Workshop on Frontiers of New Physics: Colliders and Beyond, ICTP, Trieste, Italy, June 2014,
   \href{http://indico.ictp.it/event/a13203/session/1/contribution/3/material/0/0.pdf}{{\tt http://indico.ictp.it/event/a13203/session/1/contribution/3/material/0/0.pdf}}.



\bibitem{DeSimone:2011va}
  A.~De Simone, R.~Franceschini, G.~F.~Giudice, D.~Pappadopulo and R.~Rattazzi,
  \emph{Lopsided Gauge Mediation},
  JHEP {\bf 1105} (2011) 112
     [\href{http://xxx.lanl.gov/abs/1103.6033}{{\tt arXiv:1103.6033}}].

\bibitem{Matchev:1999ft}
  K.~T.~Matchev and S.~D.~Thomas,
  \emph{Higgs and $Z$ boson signatures of SUSY},
  \emph{Phys.\ Rev.\ D} {\bf 62} (2000) 077702
   [\href{http://xxx.lanl.gov/abs/hep-ph/9908482}{{\tt hep-ph/9908482}}].
 
 \bibitem{Meade:2009qv}
  P.~Meade, M.~Reece and D.~Shih,
  \emph{Prompt Decays of General Neutralino NLSPs at the Tevatron},
  \emph{JHEP} {\bf 1005} (2010) 105
     [\href{http://xxx.lanl.gov/abs/0911.4130}{{\tt arXiv:0911.4130}}].
 
 \bibitem{Goodman:1984dc}
  M.~W.~Goodman and E.~Witten,
  \emph{Detectability of Certain Dark Matter Candidates},
  Phys.\ Rev.\ D {\bf 31} (1985) 3059.
 
 \bibitem{Hisano:2010fy}
  J.~Hisano, K.~Ishiwata and N.~Nagata,
  \emph{A complete calculation for direct detection of Wino dark matter},
  \emph{Phys.\ Lett.\ B} {\bf 690} (2010) 311
   [\href{http://xxx.lanl.gov/abs/1004.4090}{{\tt arXiv:1004.4090}}].
 
 \bibitem{Hisano:2011cs}
  J.~Hisano, K.~Ishiwata, N.~Nagata and T.~Takesako,
  \emph{Direct Detection of Electroweak-Interacting Dark Matter},
  \emph{JHEP} {\bf 1107} (2011) 005
     [\href{http://xxx.lanl.gov/abs/1104.0228}{{\tt arXiv:1104.0228}}].
 
 \bibitem{Hisano:2012wm}
  J.~Hisano, K.~Ishiwata and N.~Nagata,
  \emph{Direct Search of Dark Matter in High-Scale SUSY},
  \emph{Phys.\ Rev.\ D} {\bf 87} (2013) 035020
       [\href{http://xxx.lanl.gov/abs/1210.5985}{{\tt arXiv:1210.5985}}].
 
 \bibitem{Hill:2011be}
  R.~J.~Hill and M.~P.~Solon,
  \emph{Universal behavior in the scattering of heavy, weakly interacting dark matter on nuclear targets},
  \emph{Phys.\ Lett.\ B} {\bf 707} (2012) 539
         [\href{http://xxx.lanl.gov/abs/1111.0016}{{\tt arXiv:1111.0016}}].

\bibitem{Hill:2013hoa}
  R.~J.~Hill and M.~P.~Solon,
  \emph{WIMP-nucleon scattering with heavy WIMP effective theory},
  \emph{Phys.\ Rev.\ Lett.\ }  {\bf 112} (2014) 211602
           [\href{http://xxx.lanl.gov/abs/1309.4092}{{\tt arXiv:1309.4092}}].

\bibitem{Hill:2014yka}
  R.~J.~Hill and M.~P.~Solon,
  \emph{Standard Model anatomy of WIMP dark matter direct detection I: weak-scale matching},
 [\href{http://xxx.lanl.gov/abs/1401.3339}{{\tt arXiv:1401.3339}}].

\bibitem{Hill:2014yxa}
  R.~J.~Hill and M.~P.~Solon,
  \emph{Standard Model anatomy of WIMP dark matter direct detection II: QCD analysis and hadronic matrix elements},
   [\href{http://xxx.lanl.gov/abs/1409.8290}{{\tt arXiv:1409.8290}}].

\bibitem{Gondolo:2004sc}
  P.~Gondolo, J.~Edsjo, P.~Ullio, L.~Bergstrom, M.~Schelke and E.~A.~Baltz,
  \emph{DarkSUSY: Computing supersymmetric dark matter properties numerically},
  \emph{JCAP} {\bf 0407} (2004) 008
  [\href{http://xxx.lanl.gov/abs/astro-ph/0406204}{{\tt astro-ph/0406204}}].
  
\bibitem{ds}
  P.~Gondolo, J.~Edsjo, P.~Ullio, L.~Bergstrom, M.~Schelke, E.~A.~Baltz, T.~Bringmann and G.~Duda, \href{http://www.darksusy.org}{{\tt http://www.darksusy.org}}.

\bibitem{Hryczuk:2011tq}
  A.~Hryczuk,
  \emph{The Sommerfeld enhancement for scalar particles and application to sfermion co-annihilation regions},
  \emph{Phys.\ Lett.\ B} {\bf 699} (2011) 271
    [\href{http://xxx.lanl.gov/abs/1102.4295}{{\tt arXiv:1102.4295}}].

\bibitem{Ade:2013zuv}
  P.~A.~R.~Ade {\it et al.}  [Planck Collaboration],
  \emph{Planck 2013 results. XVI. Cosmological parameters},
  \emph{Astron.\ Astrophys.\ } (2014)
      [\href{http://xxx.lanl.gov/abs/1303.5076}{{\tt arXiv:1303.5076}}].

\bibitem{LUXLZ}
 D.~C.~Malling {\it et al.},
  \emph{After LUX: The LZ Program},
       [\href{http://xxx.lanl.gov/abs/1110.0103}{{\tt arXiv:1110.0103}}].

\bibitem{Cushman:2013zza}
  P.~Cushman, C.~Galbiati, D.~N.~McKinsey, H.~Robertson, T.~M.~P.~Tait, D.~Bauer, A.~Borgland and B.~Cabrera {\it et al.},
  \emph{Working Group Report: WIMP Dark Matter Direct Detection},
         [\href{http://xxx.lanl.gov/abs/1310.8327}{{\tt arXiv:1310.8327}}].

\bibitem{Billard:2013qya}
  J.~Billard, L.~Strigari and E.~Figueroa-Feliciano,
  \emph{Implication of neutrino backgrounds on the reach of next generation dark matter direct detection experiments},
  \emph{Phys.\ Rev.\ D} {\bf 89} (2014) 023524
         [\href{http://xxx.lanl.gov/abs/1307.5458}{{\tt arXiv:1307.5458}}].

 \bibitem{LEPSUSY}
  LEP2 SUSY WORKING GROUP, ALEPH, DELPHI, L3 and OPAL experiment webpage,
   \href{http://lepsusy.web.cern.ch/lepsusy/}{{\tt http://lepsusy.web.cern.ch/lepsusy/}}.

\bibitem{Randall:1998uk}
  L.~Randall and R.~Sundrum,
  \emph{Out of this world SUSY breaking},
  Nucl.\ Phys.\ B {\bf 557} (1999) 79
           [\href{http://xxx.lanl.gov/abs/hep-th/9810155}{{\tt hep-th/9810155}}].

\bibitem{Giudice:1998xp}
  G.~F.~Giudice, M.~A.~Luty, H.~Murayama and R.~Rattazzi,
  	\emph{Gaugino mass without singlets},
  JHEP {\bf 9812} (1998) 027
  [\href{http://xxx.lanl.gov/abs/hep-ph/9810442}{{\tt hep-ph/9810442}}].
  
  \bibitem{Bramante:2014tba}
  J.~Bramante, P.~J.~Fox, A.~Martin, B.~Ostdiek, T.~Plehn, T.~Schell and M.~Takeuchi,
  \emph{The Relic Neutralino Surface at a 100 TeV collider},
  [\href{http://xxx.lanl.gov/abs/1412.4789}{{\tt arXiv:1412.4789 [hep-ph]}}].

\end{thebibliography}
\end{document}